\documentclass[aps,pra,twocolumn,showpacs,floatfix]{revtex4-1}
\usepackage{graphicx,amsfonts,amssymb,amsmath,hyperref}
\usepackage{textcomp}
\usepackage{esint}

\newif\ifhyper
\hypertrue
\ifhyper       
\hypersetup{        
  citecolor = {green},
  colorlinks = {true}, % false
  urlcolor = {blue} % magenta
}

\begin{document}

\graphicspath{{./figures_submit/}}

%\input{/users/lptl/dupuis/publi/definition.tex}
%\input{/home/nd/publi/definition.tex}

%%%%%%%%%%%%%%%%%%%%%%%%%%%%%%%%%%%%%%%%%%%%%%
%AUTHOR'S MACRO
\def\Gamloc{\Gamma_{\rm loc}}
\def\Vloc{V_{\rm loc}}
\def\Zqp{Z_{\rm qp}}

\def\beq{\begin{equation}}
\def\eeq{\end{equation}}
\def\llbrace{\left\lbrace}
\def\rrbrace{\right\rbrace}

\newcommand{\Tr}{{\rm Tr}} 
\newcommand{\tr}{{\rm tr}} 
\newcommand{\sgn}{{\rm sgn}} 
\newcommand{\mean}[1]{\langle #1 \rangle}
\newcommand{\commu}[2]{[#1,#2]} 
\newcommand{\bra}[1]{\langle#1|}
\newcommand{\ket}[1]{|#1\rangle}
\newcommand{\braket}[2]{\langle #1|#2\rangle}
\newcommand{\dbraket}[3]{\langle #1|#2|#3\rangle}
\def\bravac{\langle{\rm vac}|}
\newcommand{\const}{{\rm const}} 

\newcommand{\ie}{i.e. }
\newcommand{\iet}{i.e.}
\newcommand{\eg}{e.g. }
\newcommand{\cc}{{\rm c.c.}} 
\newcommand{\hc}{{\rm h.c.}} 
\def\etal{{\it et al. }}

\def\chibf{\boldsymbol{\chi}}
\def\down{\downarrow}
\def\eps{\epsilon}
\def\gam{\gamma} 
\def\phibf{\boldsymbol{\phi}}
\def\lamb{\lambda}
\def\Lamb{\Lambda}
\def\sig{\sigma}
\def\sigbf{\boldsymbol{\sigma}} 

\def\SMF{S_{\rm MF}} 
\def\SRPA{S_{\rm RPA}} 
\def\Sint{S_{\rm int}} 
\def\Sloc{S_{\rm loc}} 

\def\half{\frac{1}{2}}

\def\a{{\bf a}}
\def\b{{\bf b}}
\def\e{{\bf e}}
\def\f{{\bf f}}
\def\g{{\bf g}}
\def\h{{\bf h}}
\def\k{{\bf k}}
\def\l{{\bf l}}
\def\m{{\bf m}}
\def\n{{\bf n}} 
\def\p{{\bf p}} 
\def\q{{\bf q}}
\def\r{{\bf r}}
\def\t{{\bf t}}
\def\u{{\bf u}}
\def\v{{\bf v}}
\def\x{{\bf x}}
\def\y{{\bf y}} 
\def\z{{\bf z}} 
\def\A{{\bf A}}
\def\B{{\bf B}}
\def\D{{\bf D}} 
\def\E{{\bf E}} 
\def\F{{\bf F}} 
\def\H{{\bf H}}  
\def\J{{\bf J}}
\def\K{{\bf K}} 

\def\G{{\bf G}}
\def\L{{\bf L}}
\def\M{{\bf M}}  
\def\O{{\bf O}} 
\def\P{{\bf P}} 
\def\Q{{\bf Q}} 
\def\R{{\bf R}}
\def\S{{\bf S}}
\def\nablabf{\boldsymbol{\nabla}}

\def\w{\omega}
\def\wn{\omega_n}
\def\wnu{\omega_\nu}
\def\wp{\omega_p} 
\def\dmu{{\partial_\mu}}
\def\dl{{\partial_l}}  
\def\dt{\partial_t} 
\def\tdt{\tilde\partial_t}
\def\dk{\partial_k}
\def\tdk{\tilde\partial_k}
\def\dx{\partial_x}
\def\dy{\partial_y} 
\def\dtau{{\partial_\tau}}

\def\intr{\int d^dr}  
\def\dintr{\displaystyle \int d^dr} 
\def\intrp{\int d^dr'}
\def\dinttau{\displaystyle \int_0^\beta d\tau}
\def\dinttaup{\displaystyle \int_0^\beta d\tau'}
\def\inttau{\int_0^\beta d\tau}
\def\inttaup{\int_0^\beta d\tau'}
\def\intx{\int d^{d+1}x} 
\def\inttaur{\int_0^\beta d\tau \int d^dr}
\def\intinf{\int_{-\infty}^\infty}
\def\dinttaur{\displaystyle \int_0^\beta d\tau \int d^dr}
\def\dintinf{\displaystyle \int_{-\infty}^\infty}
\def\intw{\int_{-\infty}^\infty \frac{d\w}{2\pi}}

\def\calA{{\cal A}} 
\def\calC{{\cal C}} 
\def\dt{\partial_t}
\def\calD{{\cal D}}
\def\calF{{\cal F}} 
\def\calG{{\cal G}}
\def\calH{{\cal H}}
\def\calI{{\cal I}}
\def\calJ{{\cal J}}
\def\calK{{\cal K}}
\def\calL{{\cal L}} 
\def\calN{{\cal N}}
\def\calO{{\cal O}}
\def\calP{{\cal P}}  
\def\calR{{\cal R}} 
\def\calS{{\cal S}}
\def\calT{{\cal T}}
\def\calU{{\cal U}}
\def\calY{{\cal Y}} 
\def\calZ{{\cal Z}} 

\def\calFbf{{\bf F}}

\def\tT{{\tilde T}}
\def\talpha{{\tilde\alpha}}
\def\tdelta{{\tilde\delta}}
\def\tlamb{{\tilde\lambda}}
\def\tmu{{\tilde\mu}}
\def\tphibf{{\tilde\phibf}}
\def\trho{{\tilde\rho}}
\def\tvarphibf{{\tilde\varphibf}} 

%%%%%%%%%%%%%%%%%%%%%%%%%%%%%%%%%%%%%%%%%%%%%%

\title{Thermodynamics of a Bose gas near the superfluid--Mott-insulator transition} 

\author{A. Ran\c{c}on and  N. Dupuis}
\affiliation{Laboratoire de Physique Th\'eorique de la Mati\`ere Condens\'ee, 
CNRS UMR 7600, \\ Universit\'e Pierre et Marie Curie, 4 Place Jussieu, 
75252 Paris Cedex 05, France}

\date{9 July 2012}

\begin{abstract} 
We study the thermodynamics near the generic (density-driven) superfluid--Mott-insulator transition in the three-dimensional Bose-Hubbard model using the nonperturbative renormalization-group approach. At low energy the physics is controlled by the Gaussian fixed point and becomes universal. Thermodynamic quantities can then be expressed in terms of the universal scaling functions of the dilute Bose gas universality class while the microscopic physics enters only {\it via} two nonuniversal parameters, namely the effective mass $m^*$ and the ``scattering length'' $a^*$ of the elementary excitations at the quantum critical point between the superfluid and Mott-insulating phase. A notable exception is the condensate density in the superfluid phase which is proportional to the quasi-particle weight $\Zqp$ of the elementary excitations. The universal regime is defined by $m^*a^*{}^2 T\ll 1$ and $m^*a^*{}^2|\delta\mu|\ll 1$, or equivalently $|\bar n-\bar n_c|a^*{}^3\ll 1$, where $\delta\mu=\mu-\mu_c$ is the 
chemical potential shift from the quantum critical point $(\mu=\mu_c,T=0)$ and $\bar n-\bar n_c$ the doping with respect to the commensurate density $\bar n_c$ of the $T=0$ Mott insulator. We compute $\Zqp$, $m^*$ and $a^*$ and find that they vary strongly with both the ratio $t/U$ between hopping amplitude and on-site repulsion  and the value of the (commensurate) density $\bar n_c$. Finally, we discuss the experimental observation of universality and the measurement of $\Zqp$, $m^*$ and $a^*$ in a cold atomic gas in an optical lattice.      
\end{abstract}
\pacs{05.30.Rt,05.30.Jp,67.85.-d,03.75.Hh}
\maketitle

\section{Introduction}

The low-temperature thermodynamics of a dilute ultracold Bose gas is well understood both theoretically and experimentally. The equation of state, {\it e.g.} the pressure $P(\mu,T)$ vs chemical potential and temperature, turns out to be ``universal'' to the extent that it depends only on a small number of parameters such as the mass $m$ of the bosons and their $s$-wave scattering length $a$, and is otherwise independent of other microscopic characteristics such as details of the atom-atom interaction potential. From a theoretical point of view, the thermodynamics of a dilute Bose gas is usually derived within a low-density expansion using $ma^2\mu$ (or $\bar na^3$, with $\bar n$ the mean boson density) as the expansion parameter. 

Strong correlations in an ultracold Bose gas can be achieved by loading the gas into an optical lattice. It is then possible to induce a quantum phase transition between a superfluid ground state and a Mott insulating phase by varying the strength of the lattice potential~\cite{Greiner02}. The main features of the Mott transition can be understood in the framework of the Bose-Hubbard model, which describes bosons moving in a lattice with an on-site repulsive interaction~\cite{Fisher89}. 

\begin{figure}
\centerline{\includegraphics[width=6cm,clip]{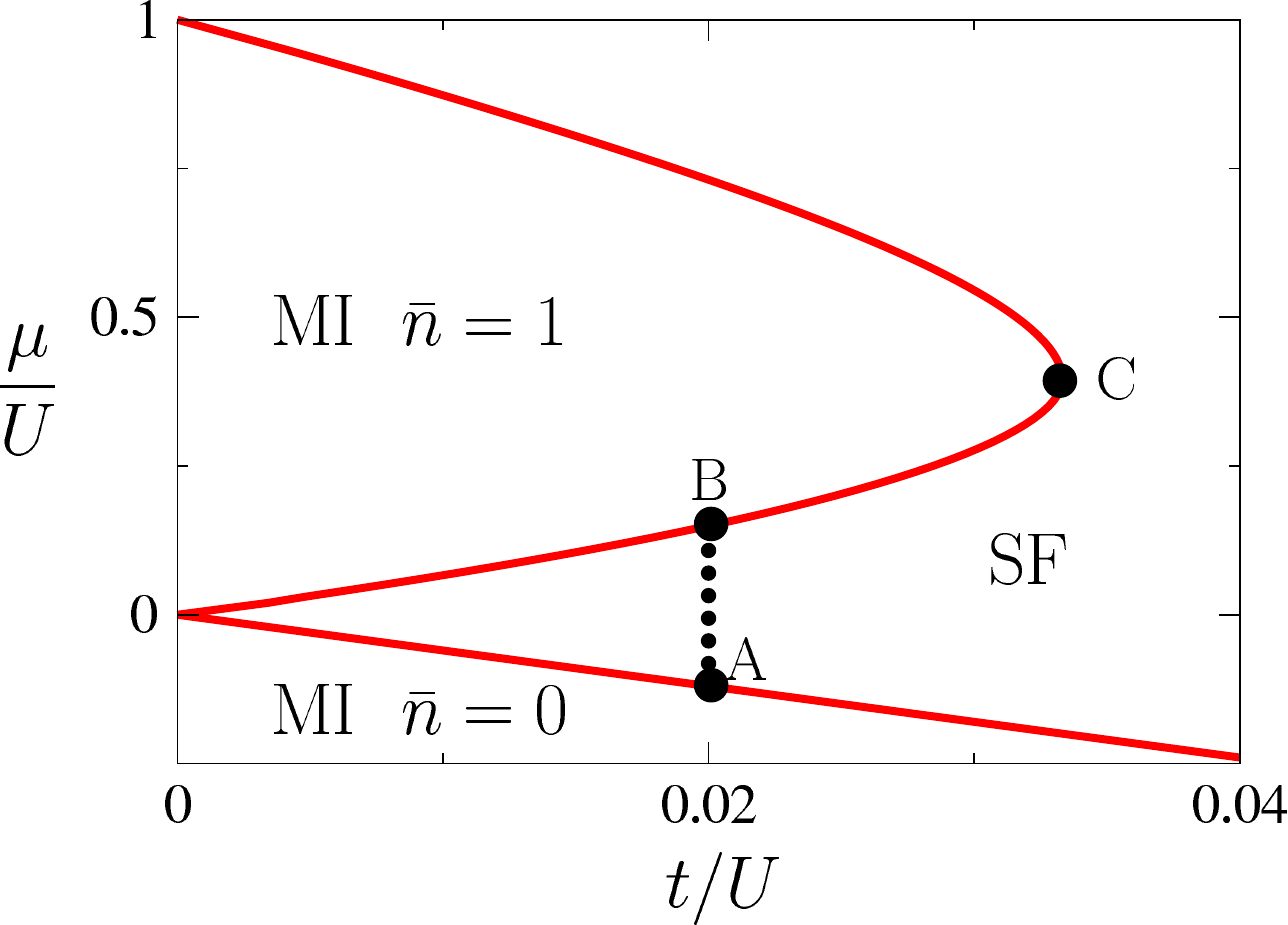}}
\caption{(Color online)  Zero-temperature phase diagram of the Bose-Hubbard model on a cubic lattice showing the Mott insulators (MI) with density $\bar n=0$ (vacuum) and $\bar n=1$, as well as the surrounding superfluid phase (SF). Point C at the tip of the Mott lobe shows the multicritical point where the transition occurs at fixed density $\bar n=1$. Away from this point, the transition is driven by a density change. The finite-temperature pressure $P(\mu_c,T)$ at point B is shown in Fig.~\ref{fig_pressure2}. The zero-temperature pressure $P(\mu,0)$, condensate density $n_0(\mu,0)$, superfluid stiffness $\rho_s(\mu,0)$, sound mode velocity $c(\mu,0)$ and superfluid transition temperature $T_c(\mu)$ along the dotted line AB are shown in Figs.~\ref{fig_pressure3}-\ref{fig_Tc}.}
\label{fig_phase_dia}
\end{figure}

In the vicinity of the Mott transition, there is no small parameter (such as density or interaction strength) that would allow us to derive the equation of state perturbatively. Nevertheless, near the generic (density-driven) Mott transition, the thermodynamics of a Bose gas turns out to be similar to that of a dilute Bose gas up to some effective parameters. The origin of this similarity can be understood as follows. By varying the chemical potential from negative to positive values in a dilute Bose gas, one induces a (zero-temperature) quantum phase transition between a state with no particles (vacuum) and a superfluid state with a finite density. This identifies the point $\mu=T=0$ as a quantum critical point (QCP). Above the upper critical dimension $d_c^+=2$, the boson-boson interaction is irrelevant (in the renormalization-group sense) and the critical behavior at the transition is mean-field like with a correlation-length exponent $\nu=1/2$ and a dynamical critical exponent $z=2$. Elementary 
excitations at the QCP are free bosons of mass $m$ and 
their mutual interaction is determined by the $s$-wave scattering length $a$ in the low-energy limit. The dependence of the equation of state of a dilute Bose gas on $m$ and $a$ only is a direct consequence of the proximity of the QCP between the superfluid phase and the vacuum: the thermodynamics is controlled by the QCP. It follows that thermodynamic quantities can be expressed in terms of universal scaling functions of $\mu/T$ and an effective temperature-dependent dimensionless interaction constant $\tilde g(T)=8\pi\sqrt{2ma^2T}$~\cite{Rancon12b}. This universal description holds in the critical regime of the QCP defined by $ma^2|\mu|$ and $ma^2T\ll 1$. 

The vacuum-superfluid transition of a dilute Bose gas and the generic Mott transition of a Bose gas in an optical lattice belong to the same universality class. Both transitions are governed by the same (Gaussian) fixed point. Elementary excitations at the QCP between the Mott insulator and the superfluid phase are quasi-particles with effective mass $m^*$ and their mutual interaction is described by an effective ``scattering length'' $a^*$. Near the QCP, thermodynamic quantities can be expressed with the universal scaling functions of the dilute Bose gas universality class and the nonuniversal parameters $m^*$ and $a^*$. This conclusion is correct  everywhere near the superfluid--Mott-insulator transition except in the close vicinity of the multicritical points where the transition takes place at fixed (commensurate) density (Fig.~\ref{fig_phase_dia}). 

In this paper, we study the thermodynamics of the three-dimensional Bose-Hubbard model using a nonperturbative renormalization-group (NPRG) approach~\cite{Rancon11a,Rancon11b,Rancon12a}. In Sec.~\ref{sec_dbg}, we derive scaling forms for various thermodynamic quantities (pressure, density, compressibility, condensate density, superfluid stiffness and superfluid transition temperature) of a dilute Bose gas and discuss the scaling functions in some limiting cases. The nonperturbative renormalization-group approach to the Bose-Hubbard model is briefly reviewed in Sec.~\ref{sec_nprg}. In Sec.~\ref{sec_bhm}, we show that the NPRG approach enables to straightforwardly identify the elementary excitations at the QCP governing the generic Mott transition, and compute their effective mass $m^*$ and effective scattering length $a^*$ as well as their spectral weight $\Zqp$. $\Zqp$, $m^*$ and $a^*$ are calculated as a function of $t/U$. We then present various thermodynamic quantities obtained from a numerical solution 
of the NPRG equations and show that near the Mott transition they satisfy the scaling behavior characteristic of the dilute Bose gas universality class, except for the condensate density which is proportional to the quasi-particle weight. The experimental implications of our results are discussed in the Conclusion.

\section{The dilute Bose gas universality class}
\label{sec_dbg}

In this section we discuss in detail the dilute Bose gas universality class. We derive scaling forms for various thermodynamic quantities and compute the corresponding universal scaling functions in some limits. 

\subsection{Universal scaling functions}
\label{subsec_univ}

Let us consider a three-dimensional Bose gas described by the (Euclidean) action 
\begin{equation} 
S = \inttau\intr \llbrace \psi^*\left(\dtau-\mu+\frac{\nablabf^2}{2m} \right) \psi 
+ \frac{g}{2} (\psi^*\psi)^2 \rrbrace , 
\label{action0}
\end{equation}
where $\psi(\r,\tau)$ is a complex field and $\tau\in[0,\beta]$ an imaginary time with $\beta=1/T$ the inverse temperature. $\mu$ denotes the chemical potential. The interaction is assumed to be local in space and the model is regularized by a ultraviolet momentum cutoff $\Lambda$. $d=3$  and we set $\hbar=k_B=1$ throughout the paper.  

The nature of the $\mu=T=0$ QCP between the vacuum and the superfluid state can be understood from a RG analysis. Since the $\mu=0$ ground state is the vacuum, there is no renormalization of the single-particle propagator and the correlation-length exponent $\nu=1/2$, the anomalous dimension $\eta=0$ while the dynamical critical exponent $z=2$. The dimensionless interaction constant $\tilde g=2mg\Lambda$ satisfies the (exact) RG equation 
\begin{equation}
s \frac{d\tilde g(s)}{ds} = -\tilde g(s) - \frac{\tilde g(s)^2}{4\pi^2}
\label{rgeq}
\end{equation}
(with $\tilde g(1)=\tilde g$), 
when fluctuation modes with momenta between $\Lambda$ and $\Lambda/s$ are integrated out (with a proper rescaling of fields, momenta and frequencies in order to restore the original value of the cutoff $\Lambda$)~\cite{Sachdev_book,Rancon12b}. From Eq.~\eqref{rgeq}, we obtain
\begin{equation}
\tilde g(s) = \frac{8\pi\Lambda a}{s} \quad \mbox{for} \quad s\gg 1 ,
\end{equation}
where 
\begin{equation}
a = \frac{mg}{4\pi + \frac{2}{\pi}mg\Lambda} 
\end{equation}
is the $s$-wave scattering length which can be calculated from the action \eqref{action0} by solving the two-body problem. $\tilde g(s)$ is thus irrelevant (it vanishes for $s\to\infty$) and the only fixed point of Eq.~\eqref{rgeq} is $\tilde g =0$ in agreement with the fact that the upper critical dimension for the vacuum-superfluid transition is $d_c^+=2$. 

There are two relevant perturbations about the Gaussian fixed point $\mu=T=\tilde g=0$: the chemical potential $\mu$ and the temperature $T$, with scaling dimensions $[\mu]=1/\nu$ and $[T]=z$. In a RG transformation, they transform as $\mu(s)=s^{1/\nu}\mu$ and $T(s)=s^zT$. In the critical regime near the QCP, the pressure satisfies~\cite{note5}
\begin{equation}
P(\mu,T) = s^{-d-z} P(s^{1/\nu}\mu,s^z T,\tilde g(s)) .
\label{Pscaling}
\end{equation}
By choosing $s\sim T^{-1/z}$ or $s\sim |\mu|^{-\nu}$ and setting $z=1/\nu=2$ (with $d=3$), we can write the pressure in the scaling form
\begin{equation}
P(\mu,T) = \left(\frac{m}{2\pi}\right)^{3/2} T^{5/2} \calF\left(\frac{\mu}{T},\tilde g(T)\right) ,
\label{pressure1}
\end{equation}
or
\begin{equation}
P(\mu,T) = \left(\frac{m}{2\pi}\right)^{3/2} \mu^{5/2} \calG\left(\frac{T}{\mu},\tilde g(\mu)\right) .
\label{pressure2}
\end{equation}
The overall factor $m^{3/2}$ comes from dimensional considerations while the factor $1/(2\pi)^{3/2}$ is introduced for convenience~\cite{Rancon12b}. 
The energy-dependent effective (dimensionless) interaction constant $\tilde g(\eps)\equiv g(s=\Lambda/\sqrt{2m|\eps|})$ is defined by 
\begin{equation}
\tilde g(\eps) = 8\pi \sqrt{2ma^2|\eps|} 
\label{gtilde}
\end{equation}
and is entirely determined by the mass $m$ of the bosons and the scattering length $a$. $\calF$ and $\calG$ and universal scaling functions characteristic of the three-dimensional dilute Bose gas universality class. Equations~(\ref{pressure1}) and (\ref{pressure2}) are valid in the critical regime near the QCP defined by $ma^2|\mu|\ll 1$ and $ma^2T\ll 1$. Note that the interaction constant $\tilde g$ is a dangerously irrelevant variable (in the RG sense) and therefore cannot be neglected: $\calF$ and $\calG$ are singular functions of $\tilde g(T)$ or $\tilde g(\mu)$. Higher-order interactions, such as three-body interactions, are not considered here since they are irrelevant and give rise to subleading contributions to the pressure. 

Equations~(\ref{pressure1}) and (\ref{pressure2}) imply scaling forms for other thermodynamic quantities. For example, the particle density and compressibility read
\begin{equation}
\begin{split}
\bar n(\mu,T) &= \frac{\partial P}{\partial\mu} = \left(\frac{m}{2\pi}\right)^{3/2} T^{3/2} \calF^{(1,0)}\left(\frac{\mu}{T},\tilde g(T)\right) , \\ 
\kappa(\mu,T) &= \frac{\partial^2 P}{\partial\mu^2} = \left(\frac{m}{2\pi}\right)^{3/2} T^{1/2} \calF^{(2,0)}\left(\frac{\mu}{T},\tilde g(T)\right) ,
\end{split}
\label{nkappa}
\end{equation}
respectively, where we use the notation $\calF^{(i,j)}(x,y)=\partial^i_x \partial^j_y \calF(x,y)$. 

For positive chemical potential, there is a superfluid transition at a temperature $T_c$. This transition corresponds to a singularity of the scaling function $\calF(x,y)$ when $x=x_c(y)$. It follows that 
\begin{equation}
\frac{\mu}{T_c} = \calH\bigl(\tilde g(T_c)\bigr) ,
\label{muTc}
\end{equation}
with $\calH$ a universal scaling function. Equation~(\ref{muTc}) implies that $ma^2T_c$ is a universal function of $ma^2\mu$. 

In the superfluid phase, using~\cite{note5}
\begin{equation}
n_0(\mu,T) = s^{-d-z+2}n_0(s^{1/\nu}\mu,s^z T,\tilde g(s)) 
\label{n0scaling} 
\end{equation}
with $s\sim |\mu|^{-\nu}$, one finds that the condensate density satisfies the scaling form 
\begin{equation}
n_0(\mu,T) = \left(\frac{m\mu}{2\pi}\right)^{3/2} \calI\left(\frac{T}{\mu},\tilde g(\mu)\right) , 
\label{n0}
\end{equation}
with $\calI$ a universal function and $\tilde g(\mu)$ defined by~(\ref{gtilde}). The superfluid density (or superfluid stiffness $\rho_s=n_s/m$) satisfies a similar scaling form, 
\begin{equation}
n_s(\mu,T) = \left(\frac{m\mu}{2\pi}\right)^{3/2} \calJ\left(\frac{T}{\mu},\tilde g(\mu)\right) .
\label{ns}
\end{equation}
Galilean invariance implies that the $T=0$ superfluid density $n_s(\mu,0)$ is equal to the fluid density $\bar n(\mu,0)$ and is therefore determined by the scaling function $\calF$ [Eq.~(\ref{nkappa})]. The sound mode velocity can be expressed in terms of the compressibility and the superfluid stiffness $\rho_s=n_s/m$ (see, e.g., Ref.~\cite{Dupuis09b}), 
\begin{equation}
c(\mu,T) = \sqrt{\frac{\rho_s(\mu,T)}{\kappa(\mu,T)}} .
\label{vel}
\end{equation}
At zero-temperature, since $n_s=\bar n$, the Bogoliubov sound mode velocity $c(\mu,0)$ is equal to the macroscopic sound velocity~\cite{Gavoret64}.

\subsection{Limiting cases} 
\label{subsec_limitcases}

For a three-dimensional Bose gas, the scaling functions can be obtained from perturbation theory (see Appendix~\ref{sec_scaling}). In this section, we discuss various limiting cases. 

\subsubsection{Dilute classical gas}

When the chemical potential is large and negative, $\mu<0$ and $|\mu|\gg T$, the system behaves as a dilute classical gas and the pressure takes the form 
\begin{equation}
P(\mu,T) = \left(\frac{m}{2\pi}\right)^{3/2} T^{5/2} e^{-|\mu|/T} , 
\label{Pdilute}
\end{equation}
which leads to \
\begin{equation}
\begin{array}{lcc}
\calF(x,y) = e^x & \mbox{for} & x<0 \mbox{\;and\;} |x|\gg 1, \\
\calG(x,y) = x^{5/2} e^{1/x} & \mbox{for} & x<0 \mbox{\;and\;} |x|\ll 1 .
\end{array}
\end{equation}

\subsubsection{$T=0$ superfluid phase}

At low temperatures and positive chemical potential, the scaling functions can be obtained from the Bogoliubov theory~\cite{Bogoliubov47,Fetter_book}. At zero temperature, 
\begin{align}
P(\mu,0) &= \frac{m\mu^2}{8\pi a} \left(1-\frac{64}{15\pi} \sqrt{ma^2\mu}\right) , \label{P1} \\
\bar n(\mu,0) &= \frac{m\mu}{4\pi a} \left(1-\frac{16}{3\pi} \sqrt{ma^2\mu}\right) , \label{n1} \\
\kappa(\mu,0) &= \frac{m}{4\pi a} \left(1-\frac{8}{\pi} \sqrt{ma^2\mu}\right) , \label{kappa1} \\
n_0(\mu,0) &= \frac{m\mu}{4\pi a} \left(1 - \frac{20}{3\pi} \sqrt{ma^2\mu} \right) , \label{n0_1}
\end{align} 
and $\rho_s(\mu,0)=n_s(\mu,0)/m=\bar n(\mu,0)/m$. The first term in these equations is usually referred to as the mean-field result and the second-one as the Lee-Huang-Yang correction~\cite{Lee57a,Lee57b}. Equations~(\ref{P1}-\ref{n0_1}) can be cast in the form~(\ref{pressure2},\ref{n0},\ref{ns}) with 
\begin{equation}
\begin{split}
\calG(0,y) &= \frac{4\pi^{3/2}}{y}\left(1 - \frac{4\sqrt{2}y}{15\pi^2} \right) , \\ 
\calI(0,y) &= \frac{8\pi^{3/2}}{y} \left(1 - \frac{5\sqrt{2}y}{12\pi^2}\right) , \\ 
\calJ(0,y) &= \frac{8\pi^{3/2}}{y} \left(1 - \frac{\sqrt{2}y}{3\pi^2}\right) . 
\end{split} 
\end{equation} 
From Eqs.~(\ref{n1},\ref{kappa1}), we deduce the expression of the sound mode velocity~(\ref{vel}), 
\begin{equation}
c(\mu,0) = \sqrt{\frac{\mu}{m}} ,
\label{vel1}
\end{equation}
to leading order in $ma^2\mu$.

\subsubsection{Quantum critical regime $\mu=0$}

At vanishing chemical potential, the condensate density vanishes and the Bogoliubov theory reproduces the free boson result 
\begin{align}
P(0,T) &= -\frac{1}{\beta} \int_\q \ln\left(1-e^{-\beta\eps_\q}\right) \nonumber \\ 
&= \zeta(5/2) \left(\frac{m}{2\pi}\right)^{3/2} T^{5/2} , 
\label{P2}
\end{align}
where $\zeta(x)$ is the Riemann zeta function and $\zeta(5/2)\simeq 1.3415$. Equation~(\ref{P2}) implies
\begin{equation}
\calF(0,y) = \zeta(5/2) \quad \mbox{for} \quad y\to 0 . 
\end{equation} 

\subsubsection{Superfluid transition} 

The Bogoliubov theory correctly describes the ground state and its elementary excitations but fails near the superfluid transition temperature $T_c$. In particular, it predicts a first-order phase transition~\cite{Baym77}. The transition temperature can nevertheless be determined from a perturbative approach in the normal phase, by considering the self-consistent one-loop correction to the self-energy (self-consistent Hartree-Fock approximation),
\begin{equation}
\mu = \frac{8\pi a}{m} \zeta(3/2) \left(\frac{mT_c}{2\pi}\right)^{3/2} ,
\label{muc}
\end{equation}
which leads to      
\begin{equation}
\calH(x) = \frac{\zeta(3/2)}{4\pi^{3/2}} x .
\label{calH} 
\end{equation}

\section{Lattice NPRG} 
\label{sec_nprg} 

In this section, we briefly review the NPRG approach to the Bose-Hubbard model defined on a cubic lattice~\cite{Rancon11a,Rancon11b}. The model is defined by the action 
\begin{align}
S = \inttau \biggl\lbrace & \sum_\r \Bigl[ \psi_\r^* (\dtau-\mu)\psi_\r + \frac{U}{2} (\psi_\r^*\psi_\r)^2 \Bigr] \nonumber \\ & - t \sum_{\mean{\r,\r'}} \left(\psi_\r^* \psi_{\r'}+\mbox{c.c.}\right) \biggr\rbrace ,
\label{action1}
\end{align}
where $\psi_\r(\tau)$ is a complex field. $\lbrace\r\rbrace$ denotes the $N$ sites of the lattice and $\mean{\r,\r'}$ nearest-neighbor sites. $U$ is the on-site repulsion and $t$ the hopping amplitude. We take the lattice spacing as the unit length throughout the paper. 

\subsection{Scale-dependent effective action}

The NPRG is implemented by considering a family of models with action $S_k=S+\Delta S_k$ indexed by a momentum scale $k$ varying between a microscopic scale $\Lambda$ down to 0~\cite{Berges02,Delamotte07}. $\Delta S_k$ is a ``regulator'' term defined by 
\begin{equation}
\Delta S_k=\inttau \sum_\q \psi^*(\q) R_k(\q) \psi(\q) ,
\end{equation}
where $\psi(\q)$ is the Fourier transform of $\psi_\r$ and the sum over $\q$ runs over the first Brillouin zone $]-\pi,\pi]^d$ of the reciprocal lattice. The cutoff function $R_k(\q)$ modifies the bare dispersion $t_\q=-2t\sum_{i=1}^d \cos q_i$ of the bosons. In the lattice scheme, $R_\Lambda(\q)$ is chosen such that the effective (bare) dispersion $t_\q+R_\Lambda(\q)$ vanishes~\cite{Machado10}. The action $S_\Lambda=S+\Delta S_\Lambda$ then corresponds to the local limit of decoupled sites (vanishing hopping amplitude).  

In practice, we choose the cutoff function
\begin{equation} 
R_k(\q) = - Z_{A,k} \eps_k \mbox{sgn}(t_\q) (1-y_\q)\Theta(1-y_\q) ,
\label{cutoff} 
\end{equation}
with $\Lambda=\sqrt{2d}$, $\eps_k=tk^2$, $y_\q=(2dt-|t_\q|)/\eps_k$ and $\Theta(x)$ the step function. The $k$-dependent constant $Z_{A,k}$ is defined below ($Z_{A,\Lambda}=1$). 
Since $R_{k=0}(\q)=0$, the action $S_{k=0}$ coincides with the action~(\ref{action1}) of the original model. For small $k$, the function $R_k(\q)$ gives a ``mass'' $\sim k^2$ to the low-energy modes $|\q|\lesssim k$ and acts as an infrared regulator as in the standard NPRG scheme~\cite{Berges02,Delamotte07}. 

The scale-dependent effective action 
\begin{align}
\Gamma_k[\phi^*,\phi] ={}& - \ln Z_k[J^*,J] + \inttau \sum_\r (J^*_\r\phi_\r+\mbox{c.c.}) \nonumber \\ & - \Delta S_k[\phi^*,\phi] 
\end{align}
is defined as a (slightly modified) Legendre transform which includes the explicit subtraction of $\Delta S_k[\phi^*,\phi]$. Here $Z_k[J^*,J]$ is the partition function obtained from the action $S+\Delta S_k$, $J_\r$ a complex external source which couples linearly to the bosonic field $\psi_\r$, and 
\begin{equation}
\phi_\r(\tau) = \mean{\psi_\r(\tau)} = \frac{\delta\ln Z_k[J^*,J]}{\delta J^*_\r(\tau)}
\end{equation}
the superfluid order parameter. The variation of the effective action with $k$ is governed by Wetterich's equation~\cite{Wetterich93}, 
\begin{equation}
\partial_k \Gamma_k[\phi^*,\phi] = \half \Tr\biggl\lbrace \partial_k R_k\left(\Gamma^{(2)}_k[\phi^*,\phi] + R_k\right)^{-1} \biggr\rbrace ,
\label{wetteq}
\end{equation}
where $\Gamma^{(2)}_k$ is the second-order functional derivative of $\Gamma_k$. In Fourier space, the trace in (\ref{wetteq}) involves a sum over momenta and frequencies as well as the two components of the complex field $\phi$. 

We are primarily interested in two quantities. The first one is the effective potential defined by
\begin{equation}
V_k(n)=\frac{1}{\beta N}\Gamma_k[\phi^*,\phi] \biggl|_{\phi\;\const} 
\end{equation}
where $\phi$ is a constant (uniform and time-independent) field. The U(1) symmetry of the action implies that $V_k(n)$ is a function of $n=|\phi|^2$. Its minimum determines the condensate density $n_{0,k}$ and the thermodynamic potential (per site) $V_{0,k}=V_k(n_{0,k})$ in the equilibrium state. 

The second quantity of interest is the two-point vertex 
\begin{equation}
\Gamma^{(2)}_{k,ij}(\r-\r',\tau-\tau';\phi) = \frac{\delta^{(2)} \Gamma[\phi]}{\delta\phi_{i\r}(\tau) \delta\phi_{j\r'}(\tau')} \biggl|_{\phi\;\const}  
\end{equation}
which determines the one-particle propagator $G_k=-\Gamma^{(2)-1}_k$ and enters the flow equation~(\ref{wetteq}). Here the indices $i,j$ refer to the real and imaginary parts of $\phi$,
\begin{equation}
\phi_\r(\tau) = \frac{1}{\sqrt{2}} \left[ \phi_{1\r}(\tau) + i\phi_{2\r}(\tau) \right] . 
\end{equation}
Because of the U(1) symmetry of the action~(\ref{action1}), the two-point vertex in a constant field takes the form~\cite{Dupuis09b} 
\begin{equation}
\Gamma_{k,ij}^{(2)}(q;\phi) = \delta_{i,j}\Gamma_{A,k}(q;n) + \phi_i\phi_j \Gamma_{B,k}(q;n) + \eps_{ij} \Gamma_{C,k}(q;n) 
\label{gam2}
\end{equation}
in Fourier space, where $q=(\q,i\w)$, $\w$ is a Matsubara frequency and $\eps_{ij}$ the antisymmetric tensor. For $q=0$, we can relate $\Gamma^{(2)}_k$ to the derivative of the effective potential, 
\begin{equation}
\Gamma^{(2)}_{k,ij}(q=0;\phi) = \frac{\partial^2 V_k(n)}{\partial\phi_i\partial\phi_j} = \delta_{i,j}V_k'(n) + \phi_i\phi_j V_k''(n), 
\end{equation}
so that
\begin{equation}
\begin{split}
\Gamma_{A,k}(q=0;n) &= V_k'(n) , \\
\Gamma_{B,k}(q=0;n) &= V_k''(n) , \\
\Gamma_{C,k}(q=0;n) &= 0 . \\
\end{split}
\end{equation}

\subsection{Initial conditions}
\label{subsec_init}

Since the action $S+\Delta S_\Lambda\equiv S_{\rm loc}$ corresponds to the local limit, the initial value of the effective action reads 
\begin{equation}
\Gamma_\Lambda[\phi^*,\phi] = \Gamloc[\phi^*,\phi] + \inttau \sum_\q \phi^*(\q) t_\q \phi(\q) ,
\label{GamLambda}
\end{equation}
where $\Gamloc[\phi^*,\phi]$ is the effective action in the local limit ($t=0$). It is not possible to compute the functional $\Gamloc[\phi^*,\phi]$ for arbitrary time-dependent fields~\cite{Rancon11b}. One can however easily obtain the effective potential $\Vloc(n)$ and the two-point vertex $\Gamloc^{(2)}$ in a time-independent field $\phi$. These quantities are sufficient to specify the initial conditions of the flow within the approximations discussed below. 

The initial effective action $\Gamma_\Lambda$ [Eq.~(\ref{GamLambda})] treats the local fluctuations exactly but includes the intersite hopping term at the mean-field level, thus reproducing the strong-coupling random-phase approximation (RPA)~\cite{Sheshadri93,Oosten01,Sengupta05,Ohashi06,Menotti08}. 

\subsection{Approximate solutions of the NPRG equations}

To solve the NPRG equations we expand the effective potential about its minimum, 
\begin{equation}
V_k(n) = V_{0,k} + \delta_k (n-n_{0,k}) + \frac{\lamb_k}{2} (n-n_{0,k})^2 , 
\end{equation}
where 
\begin{equation}
\delta_k = \frac{\partial V_k}{\partial n}\biggl|_{n_{0,k}} , \quad
\lamb_k = \frac{\partial^2 V_k}{\partial n^2}\biggl|_{n_{0,k}} .
\label{deltalamb} 
\end{equation}
$V_{0,k}$ determines the thermodynamic potential per site and in turn the pressure $P(\mu,T)=-V_{0,k=0}$. We also use  a derivative expansion where the two-point vertex $\Gamma^{(2)}_k(q) \equiv \Gamma^{(2)}_k(q;n_{0,k})$ in the equilibrium state is defined by 
\begin{equation}
\begin{split}
\Gamma_{A,k}(q) &= Z_{A,k} (t_\q+2dt) + V_{A,k} \w^2 + \delta_k , \\ 
\Gamma_{B,k}(q) &= \lamb_k, \\ 
\Gamma_{C,k}(q) &= Z_{C,k}\w .
\end{split}
\label{gamde}
\end{equation}
The initial values $Z_{A,\Lamb}$, $V_{A,\Lamb}$, $Z_{C,\Lamb}$, $\delta_\Lamb$ and $\lamb_\Lamb$ are deduced from $\Gamma^{(2)}_\Lamb$ and $V_\Lamb(n)$. The flow equations of $Z_{A,k}$, $V_{A,k}$, $Z_{C,k}$, $\delta_k$ and $\lamb_k$ are then obtained from the exact flow equation~(\ref{wetteq}) within a simplified Blaizot--M\'endez-Galain--Wschebor scheme~\cite{Blaizot06,Benitez09}. We refer to Ref.~\cite{Rancon11b} for a detailed discussion.

\subsection{Infrared behavior}
\label{subsec_de} 

The infrared behavior can be obtained from the action 
\begin{multline}
\Gamma_k[\phi^*,\phi] = \inttau \int d^3r \bigl[ \phi^*(Z_{C,k}\dtau-V_{A,k} \partial_\tau^2 \\ -Z_{A,k}t \nablabf^2 +\delta_k)\phi + \frac{\lamb_k}{2}(n-n_{0,k})^2 + V_{0,k} \bigr] .
\label{gamcont}
\end{multline} 
Since we are interested in the low-energy limit, we consider the continuum limit where $\r$ becomes a continuous variable. Higher-order (in derivative or field) terms neglected in~(\ref{gamcont}) give subleading contributions to the infrared behavior. Most physical quantities of interest can be directly deduced from Eq.~(\ref{gamcont})~\cite{Rancon11b}. The pressure is given by
\begin{equation}
P(\mu,T) = -V_{0,k=0} .
\label{pressure5}
\end{equation}
In the superfluid phase, the superfluid stiffness can be expressed as 
\begin{equation}
\rho_{s}(\mu,T) = 2t Z_{A,k=0} n_{0,k=0} 
\label{stiffness} 
\end{equation}
and the sound velocity reads
\begin{equation}
c(\mu,T) = \sqrt{\frac{\rho_s(\mu,T)}{\kappa(\mu,T)}} ,
\end{equation}
where $\kappa=\partial^2 P/\partial\mu^2$ is obtained from~(\ref{pressure5}). 

\section{Universal thermodynamics near the Mott transition}
\label{sec_bhm}

In this section, we first discuss the QCP between the superfluid phase and the Mott insulator. We show that elementary excitations are quasi-particles with spectral weight $\Zqp$, effective mass $m^*$ and effective ``scattering length'' $a^*$. $\Zqp$, $m^*$ and $a^*$ are computed as a function of $t/U$ using the NPRG equations. We then verify that near the generic Mott transition thermodynamics quantities, as obtained from the NPRG approach, can be expressed in terms of the universal scaling functions introduced in Sec.~\ref{sec_dbg}. 

\subsection{Quantum critical point}
\label{subsec_qcp} 

At the quantum critical point ($\mu=\mu_c,T=0$) between the superfluid and Mott insulating phases, the effective action $\Gamma\equiv \Gamma_{k=0}$ [Eq.~(\ref{gamcont})] takes the form 
\begin{equation}
\Gamma[\phi^*,\phi] = \inttau \int d^3r \Bigl[ \phi^*\bigl(Z_C\dtau-Z_At\nablabf^2 \bigr)\phi  
+ \frac{\lamb}{2} |\phi|^4  \Bigr]
\label{gamc}
\end{equation}
up to a constant term $\beta N V_0$ and neglecting higher-order (in field and derivative) terms. Equation~(\ref{gamc}) is valid at a generic QCP where the transition is driven by a density change. At a multicritical point, where the transition takes place at fixed (commensurate) density, $Z_C$ vanishes and the leading time-derivative term $-V_A\dtau^2$ must be included in the effective action~\cite{Rancon11a,Rancon11b}. 

From Eq.~(\ref{gamc}), we can identify the elementary excitations at the QCP. On the lower part of the transition line (for a given Mott lobe), $Z_C$ is negative and it is convenient to perform a particle-hole transformation $\phi\leftrightarrow\phi^*$ (which changes the sign of the $\dtau$ term in~(\ref{gamc})). We can then define a quasi-particle field 
\begin{equation}
\bar\phi(\r,\tau) = \sqrt{|Z_C|} \phi(\r,\tau) , 
\label{qpfield} 
\end{equation}
and rewrite the effective action as 
\begin{equation}
\Gamma[\bar\phi^*,\bar\phi] = \inttau \int d^3r \biggl[ \bar\phi^*\biggl(\dtau-\frac{\nablabf^2}{2m^*} \biggr)\bar\phi + \half \frac{4\pi a^*}{m^*} |\bar\phi|^4\biggr] ,
\label{gam1}
\end{equation}
where 
\begin{equation}
\begin{gathered}
m^* = \frac{|Z_C|}{2tZ_A} = m \frac{|Z_C|}{Z_A} , \\
a^* = \frac{m^*\lamb}{4\pi Z_C^2} , 
\end{gathered}
\label{maeff}
\end{equation}
with $m=1/2t$ the effective mass of the free bosons moving in the lattice. We deduce from Eqs.~(\ref{qpfield},\ref{gam1}) that elementary excitations are quasi-particles with mass $m^*$ and spectral weight
\begin{equation}
\Zqp = |Z_C|^{-1}. 
\end{equation}
They are particle-like if $Z_C>0$ and hole-like if $Z_C<0$. The effective interaction between two quasi-particles is determined by the effective ``scattering length'' $a^*$. 

The quantum phase transition at $\mu=-2dt$ between the superfluid phase and the vacuum (which can be seen as a Mott insulator with vanishing density) is a particular case of a superfluid--Mott-insulator transition which differs from the superfluid-vacuum transition discussed in Sec.~\ref{subsec_univ} only by the presence of the lattice. In this case, $Z_A=Z_C=1$ (the single-particle propagator is not renormalized~\cite{Sachdev_book,Rancon12b}), so that $\Zqp=1$ and $m^*=m=1/2t$. Furthermore, the interaction constant $\lamb=4\pi a/m=8\pi ta$ can be calculated analytically and related to the scattering length 
\begin{equation}
a = \frac{1}{8\pi (t/U+A)}, \quad A\simeq 0.1264  
\label{adef}
\end{equation}
of the bosons moving in the lattice~\cite{Rancon11b}, which gives $a^*=a$. 

\begin{figure}
\centerline{\includegraphics[width=4cm,clip]{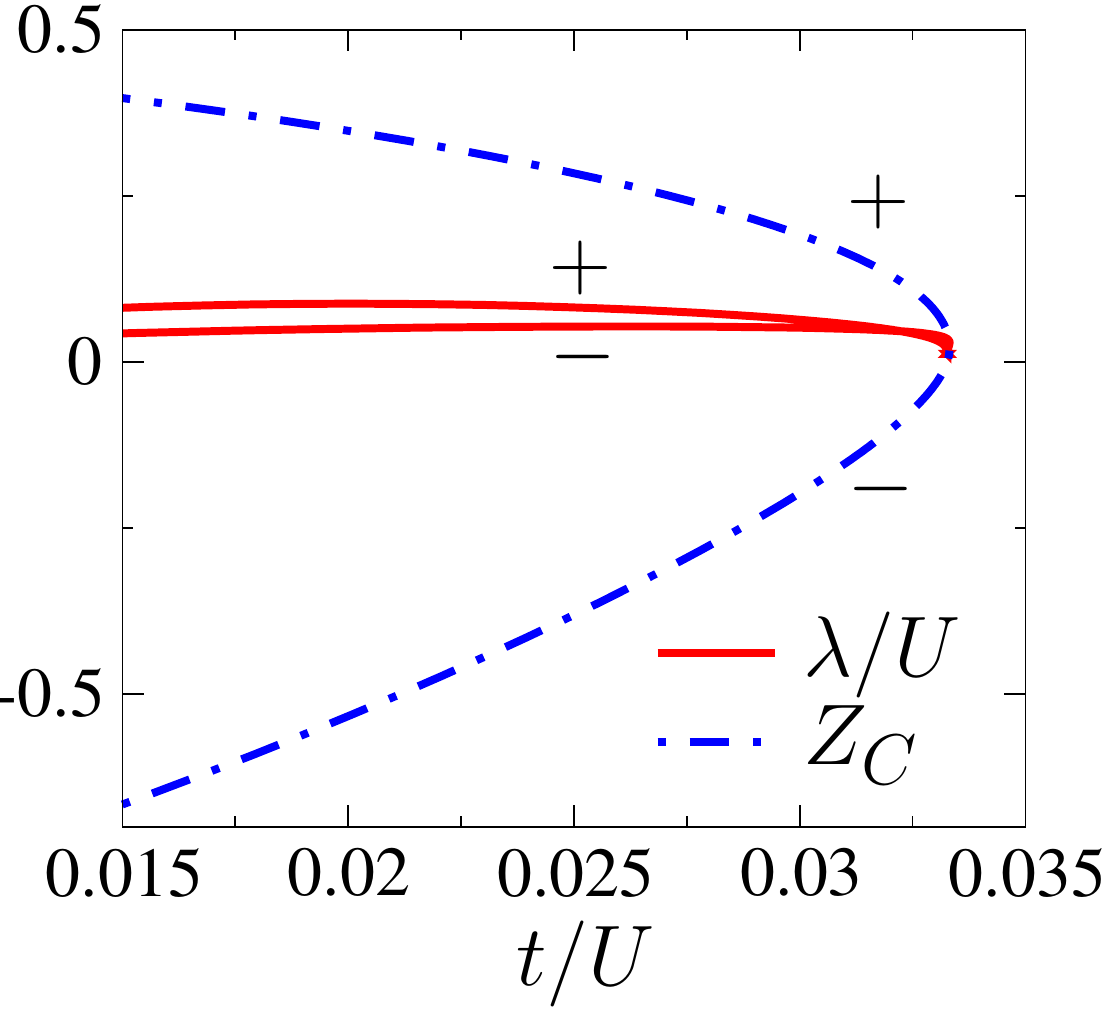}
\hspace{0.2cm}
\includegraphics[width=4cm,clip]{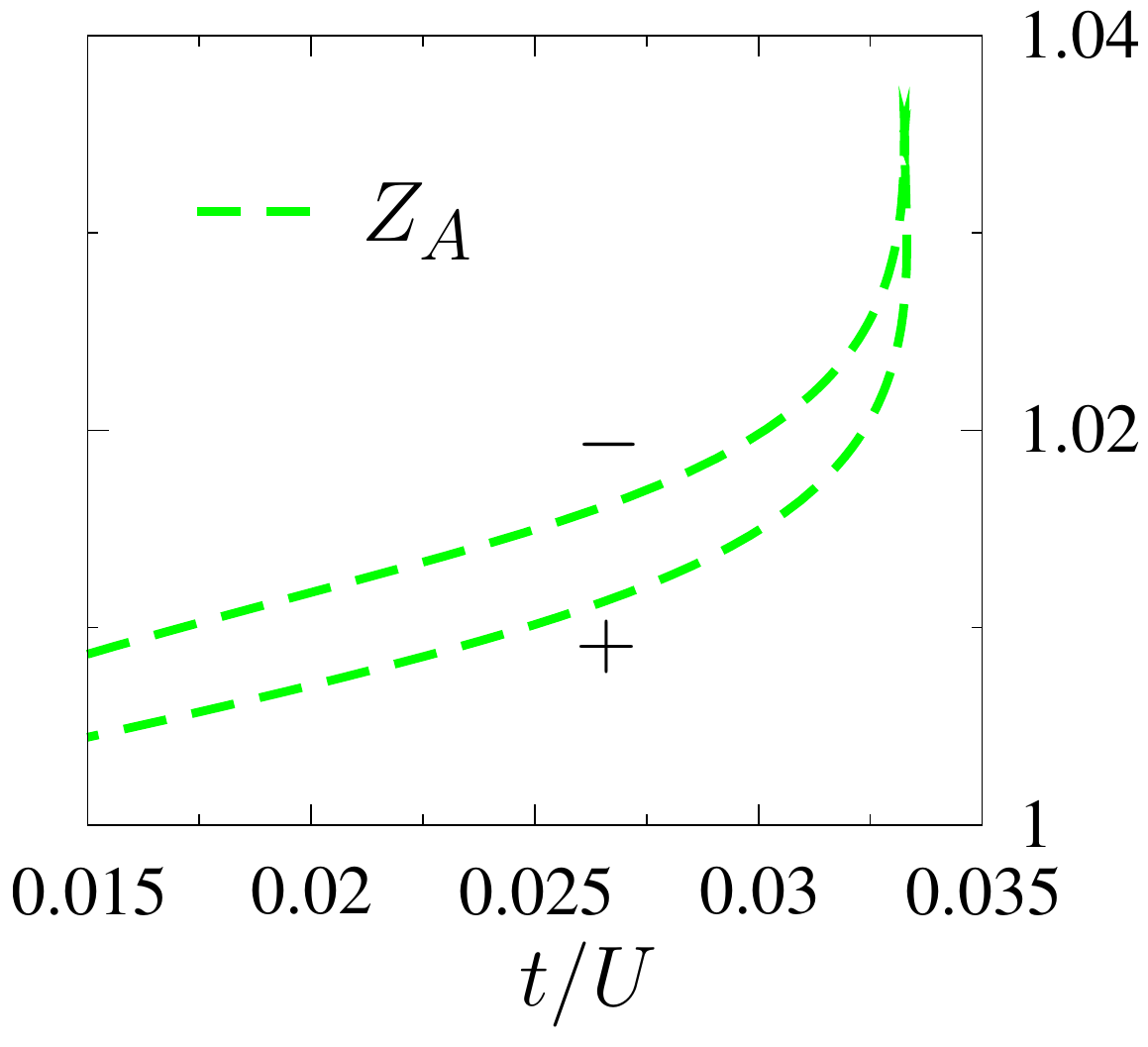}}
\caption{(Color online) $Z_C$, $\lamb$ and $Z_A$ vs $t/U$ at the QCP between the superfluid phase and the Mott insulator $\bar n=1$. The $+$ and $-$ signs refer to the upper and lower parts of the transition line.}
\label{fig_ZAZClamb}
\end{figure}
\begin{figure}
\centerline{\includegraphics[width=6.5cm,clip]{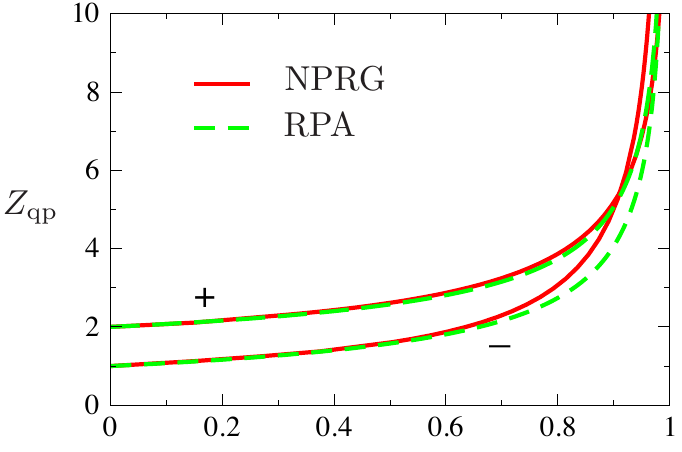}}
\centerline{\includegraphics[width=6.6cm,clip]{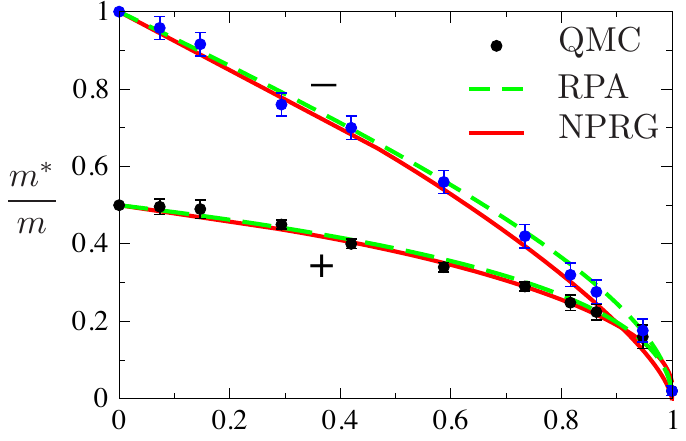}}
\centerline{\hspace{0.5cm}\includegraphics[width=6cm,clip]{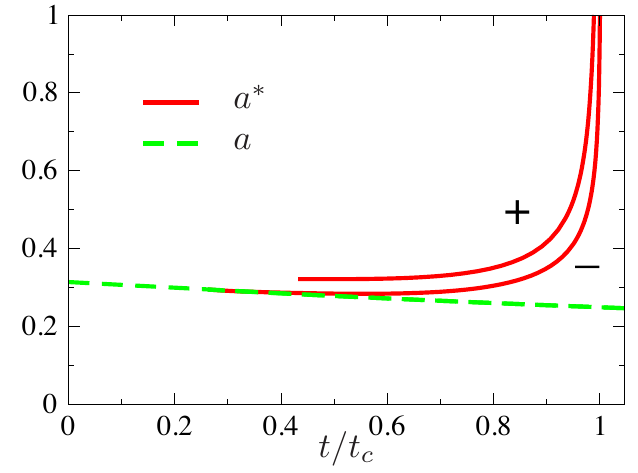}}
\caption{(Color online) Quasi-particle weight $\Zqp$, effective mass $m^*$ and scattering length $a^*$ vs $t/t_c$ at the QCP  between the superfluid phase and the Mott insulator $\bar n=1$ ($t_c$ is the value of $t$ at the tip of the Mott lobe). The QMC data are taken from Ref.~\cite{Capogrosso07}. In the bottom figure, the scattering length $a$ of the free bosons in the lattice is given by Eq.~(\ref{adef}). The $+$ and $-$ signs refer to the upper and lower parts of the transition line.}
\label{fig_ma_eff}
\end{figure}

For a generic QCP between the superfluid phase and a Mott insulating phase with nonzero density ($\bar n=1,2,3,\cdots$), the values of $\Zqp$, $m^*$ and $a^*$ can be obtained from the numerical solution of the NPRG equations. Figures~\ref{fig_ZAZClamb} and \ref{fig_ma_eff} show $Z_A,Z_C,\lamb$ and $\Zqp,m^*,a^*$ as a function of $t/U$ for the transition between the superfluid phase and the Mott insulator with density $\bar n=1$. The vanishing of $Z_C$ at the multicritical point implies that $m^*$ vanishes while $\Zqp$ and $a^*$ diverge as we approach the tip of the Mott lobe. In addition to the NPRG results, in Fig.~\ref{fig_ma_eff} we show $m^*$ obtained from quantum Monte Carlo (QMC) simulations~\cite{Capogrosso07} as well as $m^*$ and $\Zqp$ obtained from the strong-coupling random-phase approximation (RPA) (see Appendix~\ref{sec_rpa}). The RPA value for the hopping amplitude at the tip of the Mott lobe, $t_c/U\simeq 0.286$, is far away from the QMC ($t_c/U=0.034083 $)~\cite{Capogrosso07} or NPRG ($t_c/
U=0.0339$) results~\cite{note6}. Nevertheless the RPA predictions for the quasi-particle weight $\Zqp$ and the effective mass $m^*$, when plotted as a function of $t/t_c$, are in good agreement with the NPRG and QMC results (Fig.~\ref{fig_ma_eff}). As expected, the results for the lower part of the transition line are trivial in the limit $t\to 0$: $\Zqp=m^*/m=a^*/a=1$ (they are simply obtained by considering the motion of a hole in a Mott insulator with one boson per site). 

\begin{figure}
\centerline{\includegraphics[width=6.5cm,clip]{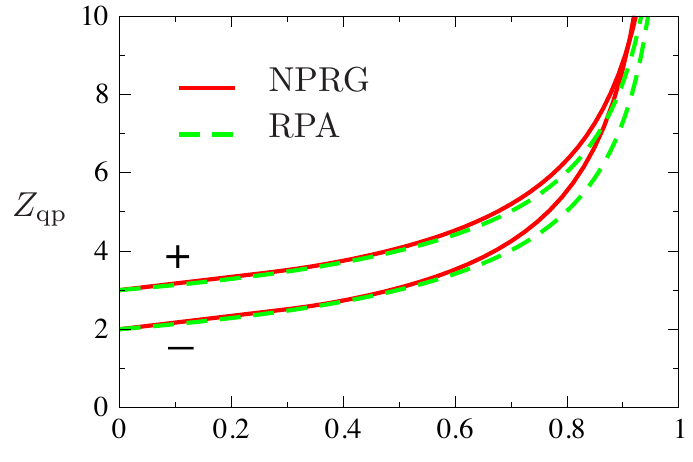}}
\centerline{\includegraphics[width=6.6cm,clip]{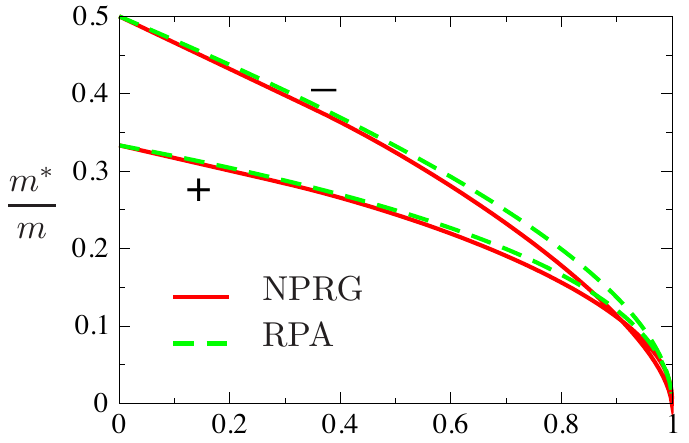}}
\centerline{\hspace{0.8cm}\includegraphics[width=5.9cm,clip]{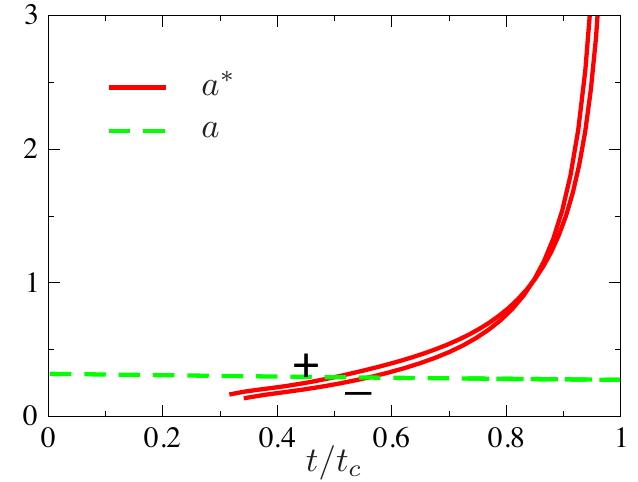}}
\caption{(Color online) Same as Fig.~\ref{fig_ma_eff} but for the transition from the superfluid phase to the Mott insulator with density $\bar n=2$.}
\label{fig_ma_eff_lobe2}
\end{figure}

In Fig.~\ref{fig_ma_eff_lobe2} we show $\Zqp$, $m^*$ and $a^*$ as a function of $t/t_c$ for the transition between the superfluid phase and the Mott insulator with density $\bar n=2$. The results are similar to the case of the transition to the first Mott lobe ($\bar n=1$) but the behavior for $t\to 0$ is different. The limiting values of the effective mass and quasi-particle weight are given by the strong-coupling RPA (see Appendix~\ref{sec_rpa}), 
\begin{equation}
\lim_{t/U\to 0} \Zqp = \lim_{t/U\to 0} \frac{m}{m^*} = \llbrace 
\begin{array}{lc}
\bar n & \mbox{(lower branch)} , \\
\bar n+1 & \mbox{(upper branch)} ,
\end{array}
\right. 
\label{tzero}
\end{equation}
where $\bar n$ denotes the (commensurate) density of the Mott insulator and we distinguish between the upper ($\mu\simeq U\bar n$) and lower ($\mu\simeq U(\bar n-1)$) branches of the transition line. 

Since $m^*/m$ and $a^*/a$ are typically of order one (except close to the Mott lobe tip), the characteristic energy scale $1/m^*a^*{}^2$ below which the physics is universal is roughly set by the hopping amplitude $t$. As we approach the tip of the Mott lobe (point C in Fig.~\ref{fig_phase_dia}), $m^*a^*{}^2$ diverges and the energy scale $1/m^*a^*{}^2$ vanishes. The low-energy physics is then controlled by the multicritical point.

\subsection{Universal thermodynamics}
\label{subsec_univ_mott}

Since the superfluid--Mott-insulator transition belongs to the dilute Bose gas universality class, near the QCP the thermodynamics can be expressed in terms of the universal scaling functions introduced in Sec.~\ref{sec_dbg} as well as the nonuniversal parameters $m^*$ and $a^*$. To ensure that there is no other nonuniversal parameter, we must verify that the chemical potential (or, more precisely, $\delta\mu=\mu-\mu_c$) couples to the elementary excitations with no additional renormalization. Slightly away from the QCP, the shift $\delta\mu=\mu-\mu_c$ in chemical potential implies a change 
\begin{equation}
\delta S=-\delta\mu \inttau \sum_\r \psi^*_\r\psi_\r
\end{equation} 
in the action. To lowest order in $\delta\mu$, $\delta S$ induces a correction 
\begin{equation}
\delta\Gamma[\phi^*,\phi] = -  Z_\mu \delta\mu\inttau \int d^3r \phi^*\phi 
\end{equation}
to the effective action~(\ref{gamc}) at the QCP, where $Z_\mu$ is a renormalization factor. Using the Ward identity $Z_\mu=Z_C$ (see Appendix~\ref{sec_Zmu}), we obtain 
\begin{align}
\delta\Gamma[\bar\phi^*,\bar\phi] &= - Z_\mu \Zqp \delta\mu \inttau \int d^3r \bar\phi^*\bar\phi \nonumber \\ &= - \sgn(Z_C) \delta \mu \inttau \int d^3r \bar\phi^*\bar\phi .
\end{align}
We conclude that $\sgn(Z_C)\delta\mu$ acts as a chemical potential for the elementary excitations at the QCP~\cite{note7}. This implies that $\pm\delta\mu/T$ will enter scaling functions with no additional scale factor. This result agrees with general considerations on the scaling of conserved densities near a continuous quantum phase transition~\cite{Sachdev94a}. 

Near the superfluid--Mott-insulator transition we can therefore write the pressure as 
\begin{equation}
P(\mu,T) = P_c + \bar n_c \delta\mu + \left(\frac{m^*}{2\pi}\right)^{3/2} T^{5/2} \calF \left(\pm \frac{\delta\mu}{T}, \tilde g(T) \right) ,
\label{pressure3}
\end{equation}
or
\begin{equation}
P(\mu,T) = P_c + \bar n_c \delta\mu + \left(\frac{m^*}{2\pi}\right)^{3/2} |\delta\mu|^{5/2} \calG \left(\pm \frac{T}{\delta\mu}, \tilde g(\delta\mu) \right) 
\label{pressure3a}
\end{equation}
where 
\begin{equation}
\tilde g(\eps) =  8\pi \sqrt{2m^*a^*{}^2|\eps|} 
\end{equation}
is obtained from Eq.~(\ref{gtilde}) by replacing $m$ and $a$ by $m^*$ and $a^*$. The $+/-$ sign in Eqs.~(\ref{pressure3},\ref{pressure3a}) corresponds to particle/hole doping of the Mott insulator ({\it i.e.} the upper/lower part of the transition line). Universality arguments imply that the singular part of the pressure can be expressed in terms of the scaling function $\calF$ but do not allows us to determine the regular part. To obtain the latter, we note that the compressibility $\kappa=\partial^2P/\partial\mu^2$ vanishes in the $T=0$ Mott insulator and has therefore no regular part, 
\begin{equation}
\kappa(\mu,T) = \left(\frac{m^*}{2\pi}\right)^{3/2} T^{1/2} \calF^{(2,0)}\left(\pm\frac{\delta\mu}{T},\tilde g(T)\right) 
\end{equation}
(see Eq.~(\ref{nkappa})). Integrating this equation twice with respect to $\mu$ yields Eq.~(\ref{pressure3}) with $P_c$ the value of the pressure at the QCP and $\bar n_c$ the density at the QCP. 

\begin{figure}
\centerline{\includegraphics[width=6cm]{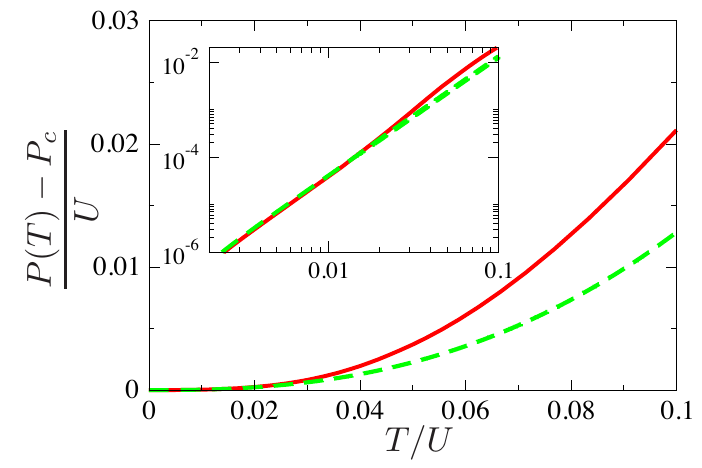}}
\vspace{-0.25cm}
\caption{(Color online) Pressure $P(\mu_c,T)$ vs temperature $T$ [$t/U=0.02$ and $\mu_c/U\simeq 0.15=$ (point B in Fig.~\ref{fig_phase_dia})]. The dashed (green) line corresponds to Eq.~(\ref{Pmuc}). The inset shows a log-log plot and the $T^{5/2}$ dependence of $P(\mu_c,T)-P_c$.}
\label{fig_pressure2}
\centerline{\includegraphics[width=5.5cm]{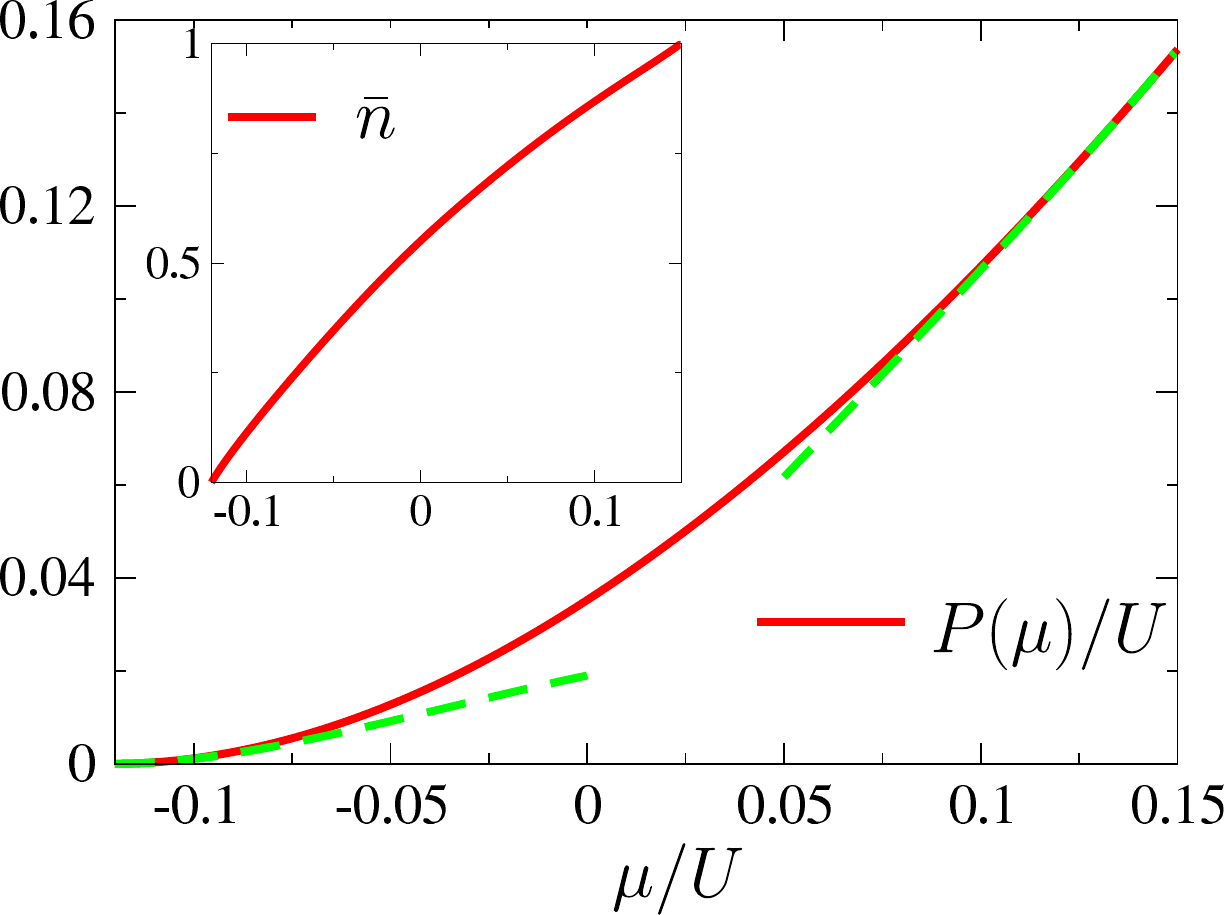}}
\vspace{0.25cm}
\centerline{\includegraphics[width=3.9cm]{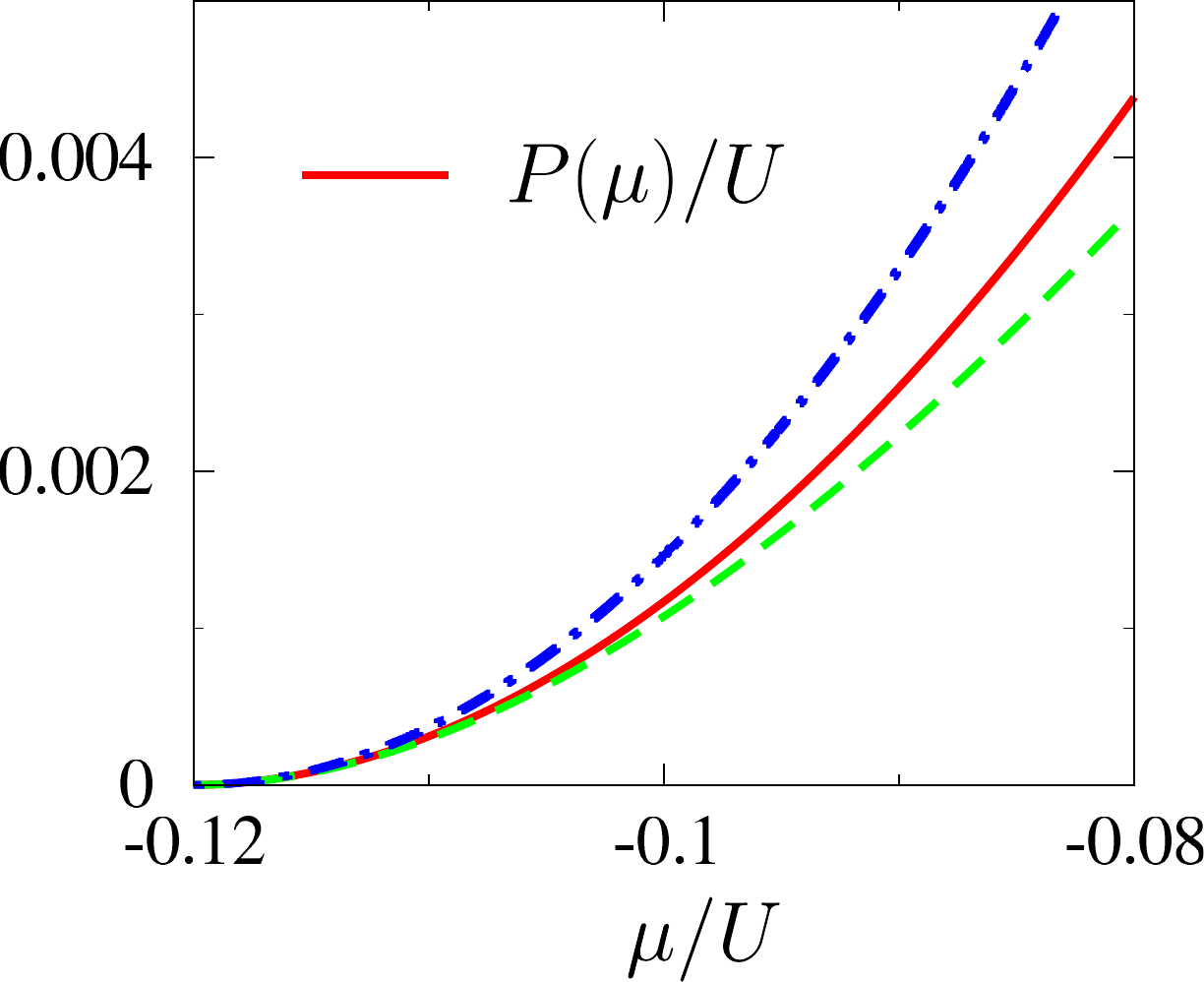}
\includegraphics[width=4.1cm]{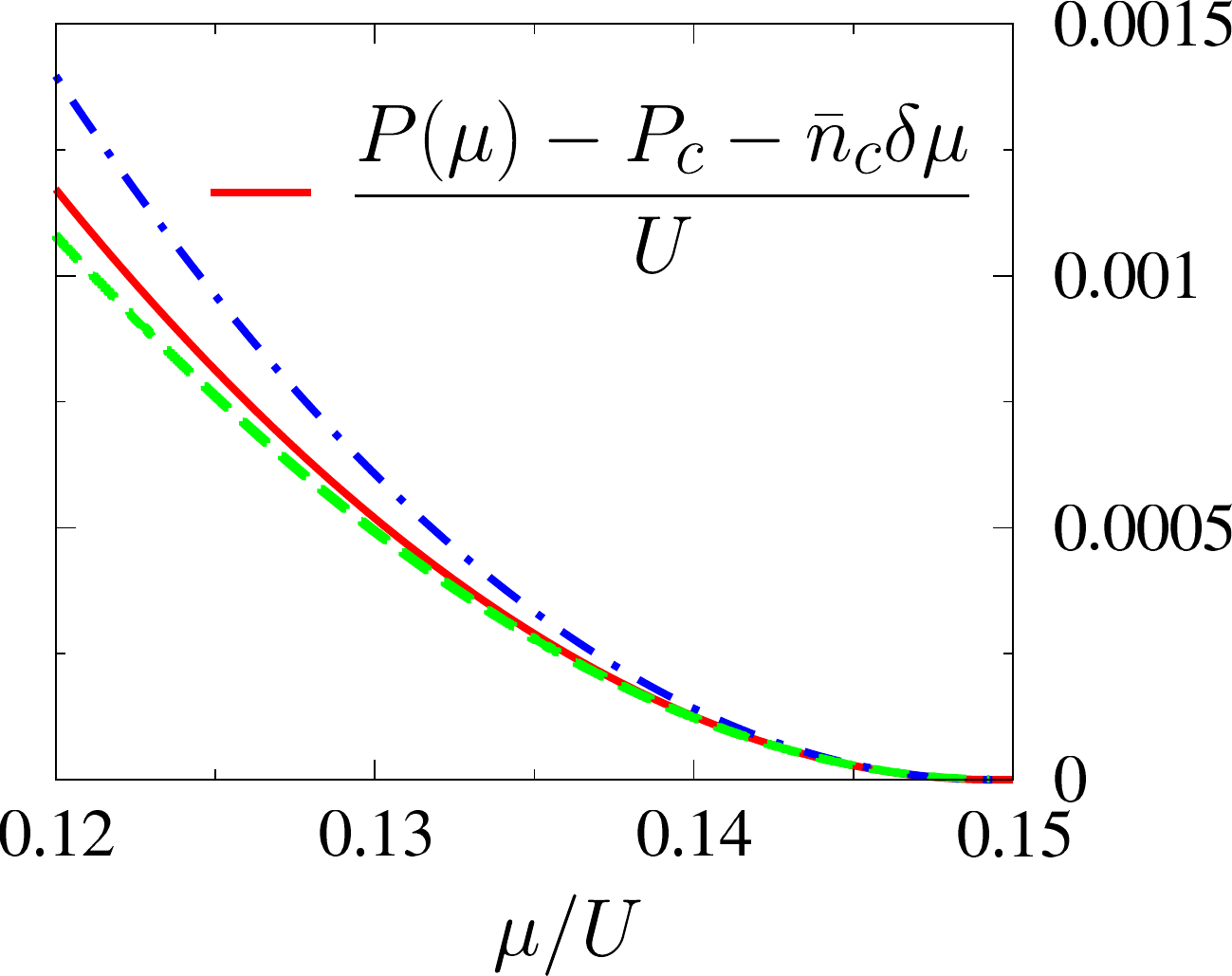}}
\caption{(Color online) Zero-temperature pressure $P(\mu,0)$ vs chemical potential $\mu$ along the dotted line in Fig.~\ref{fig_phase_dia} (the inset shows the density $\bar n=\partial P/\partial\mu$). The bottom figures show the behavior near the Mott insulating phases $\bar n=0$ and $\bar n=1$. The dashed (green) line corresponds to Eq.~(\ref{pressure4}) and the dash-dotted (blue) one to the ``mean-field'' result $P=P_c+\bar n_c\delta\mu+(\delta\mu^2)m^*/8\pi a^*$.}
\label{fig_pressure3}
\end{figure}

The results obtained from a numerical solution of the NPRG equations show that near the superfluid--Mott-insulator transition the pressure can be expressed in terms of the universal scaling function $\calF$ discussed in Sec.~\ref{sec_dbg}. In the dilute classical regime, $\sgn(Z_C)\delta\mu<0$ and $|\delta\mu|\gg T$, our results are compatible with the expected result
\begin{equation}
P(\mu,T) = P_c + \bar n_c \delta\mu + \left(\frac{m^*}{2\pi}\right)^{3/2} T^{5/2} e^{-|\delta\mu|/T}  
\end{equation}
(see Eq.~(\ref{Pdilute})). However, a precise comparison is prevented by numerical difficulties due to the extremely small values of the pressure in this regime. The finite-temperature pressure at point B in Fig.~\ref{fig_phase_dia} is shown in Fig.~\ref{fig_pressure2}. We find a very good agreement with 
\begin{equation}
P(\mu_c,T) = P_c + \zeta(5/2) \left(\frac{m^*}{2\pi}\right)^{3/2} T^{5/2} 
\label{Pmuc}
\end{equation}
below a crossover temperature scale $T\sim 2t$. Figure~\ref{fig_pressure3} shows the $T=0$ pressure at fixed $t/U$ and for a density $\bar n$ varying between 0 and 1 (see the dotted line in Fig.~\ref{fig_phase_dia}). Near the Mott insulating phases ($\bar n\simeq 0$ or $\bar n\simeq 1$), for $|\delta\mu|\lesssim t$, there is a very good agreement between the NPRG result and the universal form 
\begin{multline}
P(\mu,0) = P_c + \bar n_{c}\delta\mu \\  
+\frac{m^*(\delta\mu)^2}{8\pi a^*} \left( 1 - \frac{64}{15\pi} \sqrt{m^*a^{*2}|\delta\mu|} \right) .
\label{pressure4}
\end{multline}
It should be noted that the agreement is better with the ``Lee-Huang-Yang correction'' (last term of (\ref{pressure4})) than without. Differentiating~(\ref{pressure4}) with respect to $\mu$, we obtain 
\begin{align}
\bar n(\mu,0) &= \bar n_c + \frac{m^*\delta\mu}{4\pi a^*}\left(1 - \frac{16}{3\pi}\sqrt{m^*a^*{}^2|\delta\mu|} \right) , \label{dnbar} \\
\kappa(\mu,0) &= \frac{m^*}{4\pi a^*}\left(1 - \frac{8}{\pi}\sqrt{m^*a^*{}^2|\delta\mu|} \right) .
\label{kappa2}
\end{align}

\begin{figure}
\centerline{\includegraphics[width=7cm]{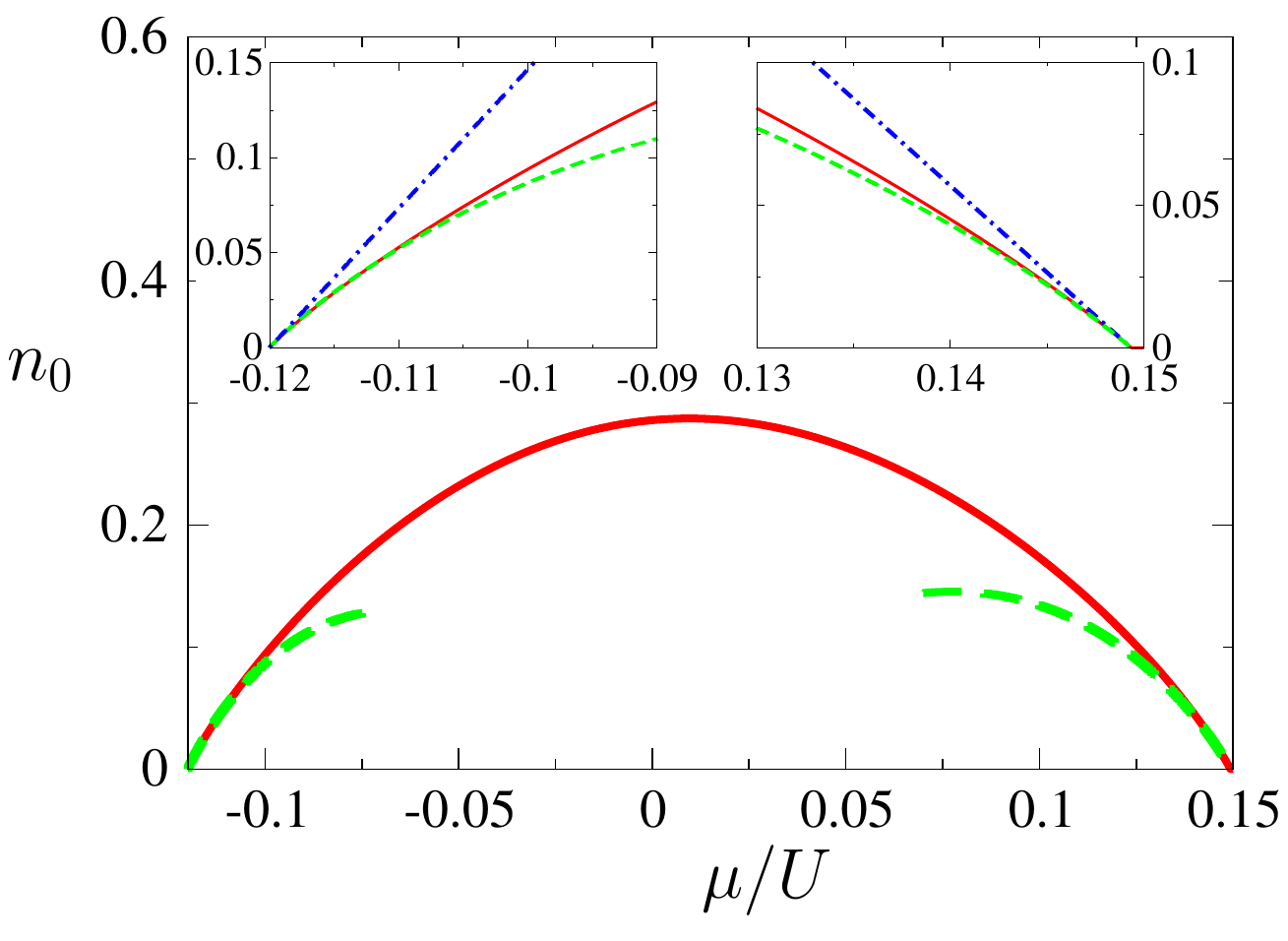}}
\vspace{-0.25cm}
\caption{(Color online) Condensate density $n_0(\mu,0)$ vs $\mu$ along the dotted line in Fig.~\ref{fig_phase_dia}. The insets show the behavior near the Mott insulating phases $\bar n=0$ and $\bar n=1$. The dashed (green) line corresponds to Eq.~(\ref{n0_2}) and the dash-dotted (blue) one to the ``mean-field'' result $n_0=\Zqp m^*|\delta\mu|/4\pi a^*$.}
\label{fig_n0}
\centerline{\includegraphics[width=7cm]{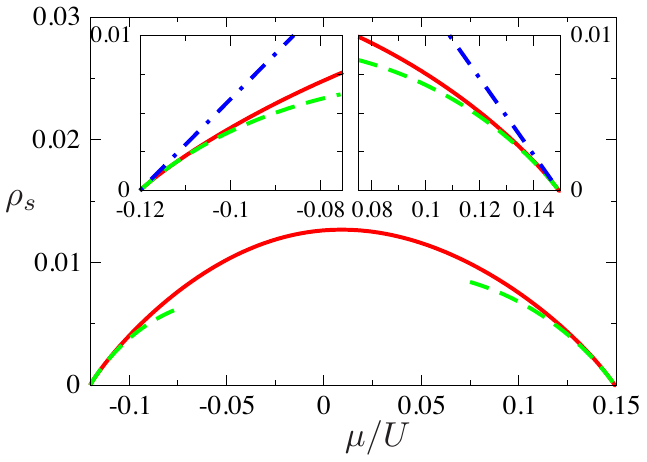}}
\caption{(Color online) Superfluid stiffness $\rho_s(\mu,0)$ vs $\mu$ along the dotted line in Fig.~\ref{fig_phase_dia}. The insets show the behavior near the Mott insulating phases $\bar n=0$ and $\bar n=1$. The dashed (green) line corresponds to Eq.~(\ref{ns_1}) and the dash-dotted (blue) one to the ``mean-field'' result $|\delta\mu|/4\pi a^*$.}
\label{fig_ns}
\end{figure}

The condensate density $n_0(\mu,T)$ in the superfluid phase can be expressed in terms of the scaling function $\calI$ [Eq.~(\ref{n0})]. However, since only the coherent part of the excitations ({\it i.e.} the quasi-particles) condenses, Eq.~(\ref{n0}) can be used for the condensate density $|\bar\phi|^2$ of the quasi-particles while $|\phi|^2=\Zqp|\bar\phi|^2$, which leads to 
\begin{equation}
n_0(\mu,T) = \Zqp \left(\frac{m^*|\delta\mu|}{2\pi}\right)^{3/2} \calI\left(\frac{T}{|\delta\mu|},\tilde g(\delta\mu)\right) 
\label{n0_3}
\end{equation} 
near the superfluid--Mott-insulator transition. The fact that $n_0(\mu,T)$, contrary to other physical quantities discussed in this section, depends on the quasi-particle weight can be understood by noting that $\phi$ is not invariant in the local gauge transformation~(\ref{localgauge}) and therefore not ``protected'' by the Ward identity~(\ref{wardid}). At zero temperature,  
\begin{equation}
n_0(\mu,0) = \Zqp \frac{m^*|\delta\mu|}{4\pi a^*} \left ( 1 - \frac{20}{3\pi} \sqrt{m^*a^{*2}|\delta\mu|} \right) ,
\label{n0_2}
\end{equation}
where $m^*|\delta\mu|/4\pi a^*\simeq |\bar n-\bar n_c|$ is the density of excess particles (or holes) with respect to the commensurate density $\bar n_c$ of the Mott insulator. The $T=0$ condensate density along the dotted line in Fig.~\ref{fig_phase_dia} is shown in Fig.~\ref{fig_n0}. Near the Mott insulating phases $\bar n_c=0$ and $\bar n_c=1$, we find a very good agreement with Eq.~(\ref{n0_2}). 

The superfluid stiffness can be expressed using the scaling function $\calJ$ [Eq.~(\ref{ns})], 
\begin{equation}
\rho_s(\mu,T) = \sqrt{m^*} \left(\frac{|\delta\mu|}{2\pi}\right)^{3/2} \calJ\left(\frac{T}{|\delta\mu|},\tilde g(\delta\mu)\right) .
\end{equation}
Using the results of Sec.~(\ref{subsec_limitcases}), we obtain 
\begin{align}
\rho_s(\mu,0) &= \frac{|\delta\mu|}{4\pi a^*}\left(1 - \frac{16}{3\pi}\sqrt{m^*a^*{}^2|\delta\mu|} \right) \nonumber \\  
&= \frac{|\bar n(\mu,0)-\bar n_c|}{m^*} ,
\label{ns_1}
\end{align} 
again in very good agreement with the NPRG approach (Fig.~\ref{fig_ns}). $\rho_s$, contrary to $n_0$, is independent of the quasi-particle weight $\Zqp$. This follows from the fact that the superfluid stiffness can be related to the current-current density correlation function~\cite{[{See, {\it e.g.}, }]Forster_book}, which is a gauge-invariant quantity. As a result, the ratio between condensate density and superfluid stiffness,  
\begin{equation}
\frac{n_0(\mu,0)}{m^*\rho_s(\mu,0)} = \Zqp 
\end{equation}
(to leading order in $m^*a^*{}^2|\delta\mu|$), explicitely depends on the quasi-particle weight while it is equal to unity in a dilute Bose gas. 

\begin{figure}
\centerline{\includegraphics[width=6cm]{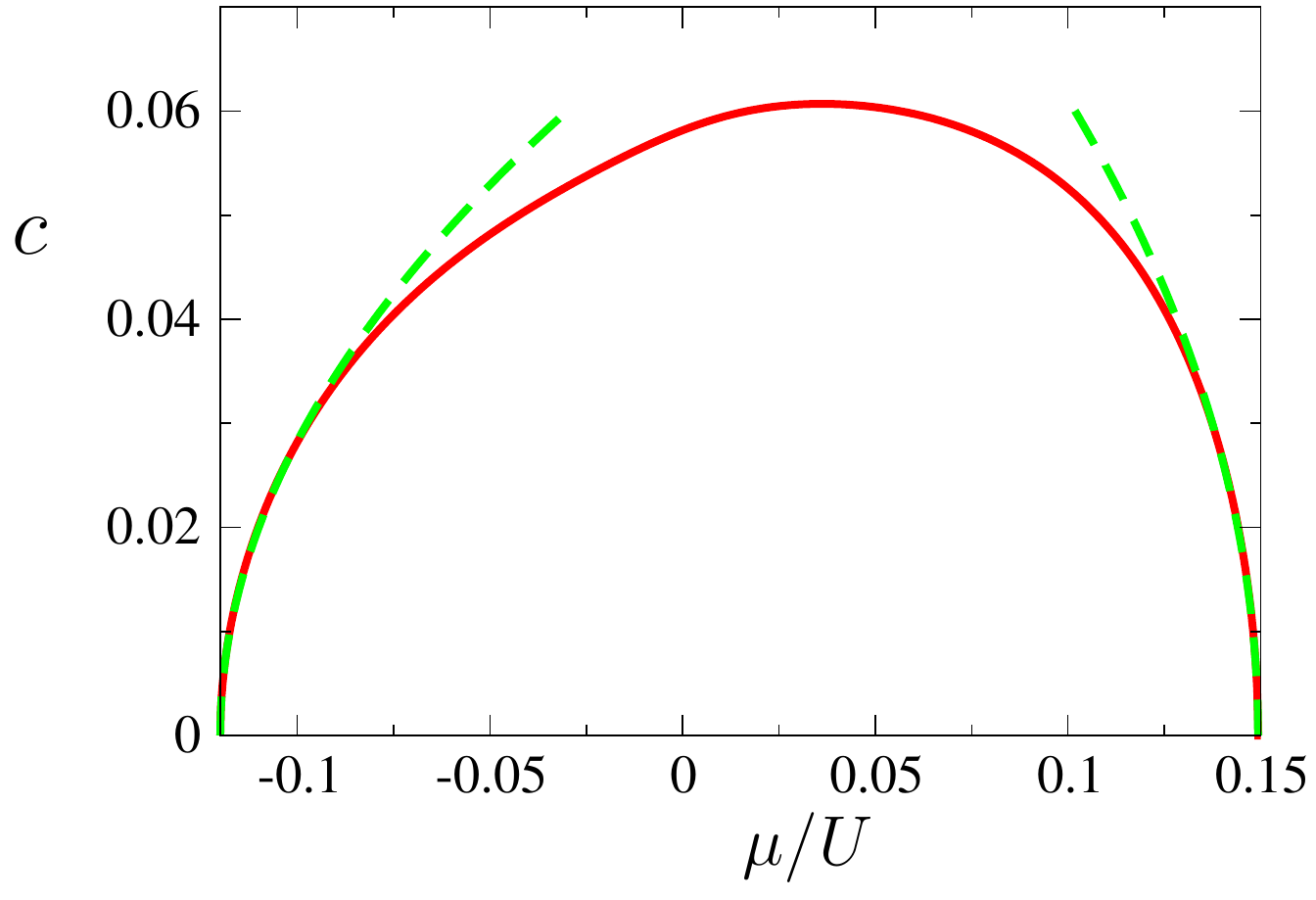}}
\vspace{-0.25cm}
\caption{(Color online) Sound mode velocity $c(\mu,0)$ along the dotted line in Fig.~\ref{fig_phase_dia}. The dashed (green) line shows the expected result near the Mott transition [Eq.~(\ref{vel3})].}
\label{fig_velocity}
\end{figure}

From Eqs.~(\ref{kappa2}) and (\ref{ns_1}), we deduce 
\begin{equation}
c(\mu,0) = \sqrt{\frac{\rho_s(\mu,0)}{\kappa(\mu,0)}} \simeq \sqrt{\frac{|\delta\mu|}{m^*}} 
\label{vel3}
\end{equation}
to leading order in $m^*a^*{}^2|\delta\mu|$. As the condensate density $n_0$ and the superfluid stiffness $\rho_s$, $c$ is related to the density of excess particles (or holes) $|\bar n-\bar n_c|$ rather than the full density $\bar n$. The NPRG results show that Eq.~(\ref{vel3}) is very well satisfied near the Mott transition (Fig.~\ref{fig_velocity}). 

\begin{figure}
\centerline{\includegraphics[width=7cm]{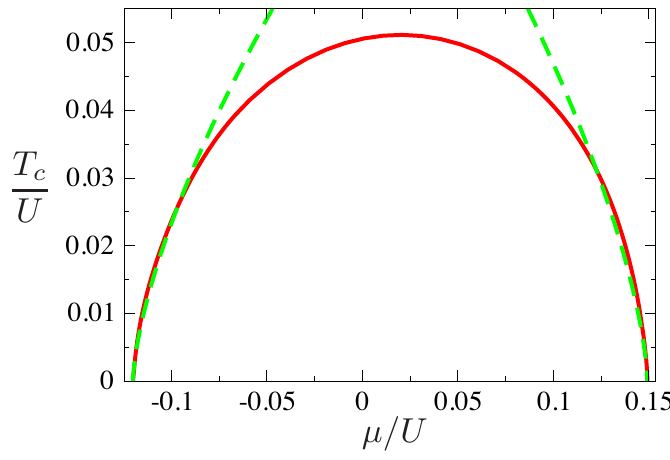}}
\vspace{-0.25cm}
\caption{(Color online) Superfluid transition temperature $T_c$ vs $\mu$ along the dotted line in Fig.~\ref{fig_phase_dia}. The dashed (green) line shows the expected result near the $T=0$ Mott transition [Eq.~(\ref{Tc})].}
\label{fig_Tc}
\end{figure}

The superfluid transition temperature is determined by the scaling function $\calH$, 
\begin{equation}
\frac{|\delta\mu|}{T_c} = \calH\bigl(\tilde g(T_c)\bigr) ,
\end{equation}
where $\calH$ is the universal scaling function introduced in Sec.~\ref{subsec_univ}. Using~(\ref{calH}), we then obtain 
\begin{equation}
T_c = \frac{2\pi}{m^*}  \left(\frac{m^*|\delta\mu|}{8\pi \zeta(3/2)a^*}\right)^{2/3} = \frac{2\pi}{m^*}  \left(\frac{|\bar n-\bar n_c|}{2\zeta(3/2)}\right)^{2/3}
\label{Tc} 
\end{equation}
(to leading order in $m^*a^*{}^2|\delta\mu|$) near the $T=0$ Mott transition, in very good agreement with the NPRG results (Fig.~\ref{fig_Tc}).  

\subsection{RG flows and approach to universality}

\begin{figure}
\centerline{\hspace{-0.52cm}
\includegraphics[width=6.3cm,clip]{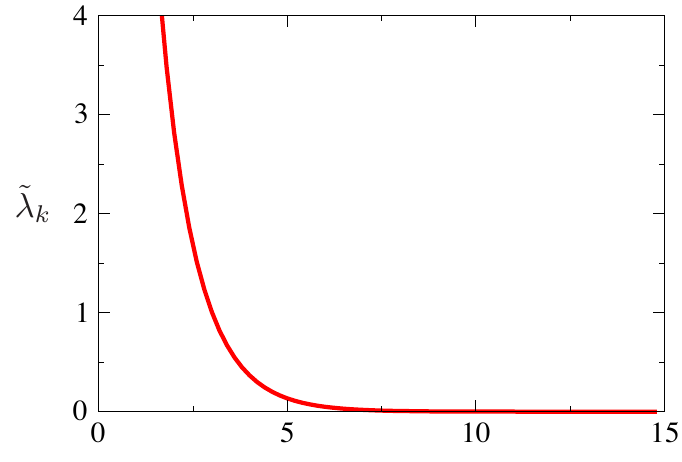}}
\centerline{\includegraphics[width=6cm,clip]{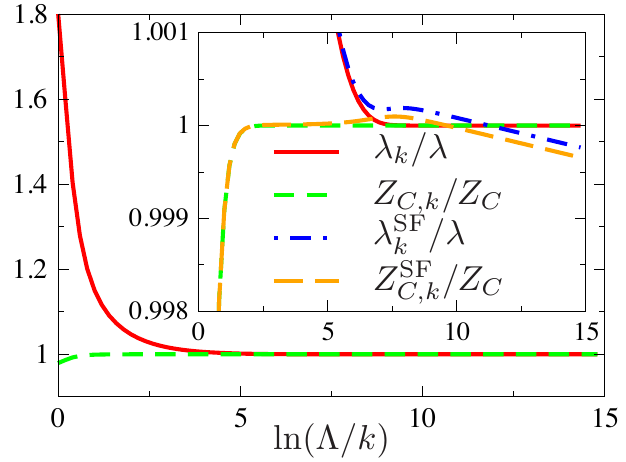}}
\caption{RG flows of $\tlamb_k$ ($\tlamb_\Lambda\simeq 37$), $\lamb_k$ and $Z_{C,k}$ at the QCP $(t/U=0.02,\mu_c/U\simeq 0.15,T=0)$ (point B in Fig.~\ref{fig_phase_dia}) and in the nearby superfluid phase $\delta\mu/U=-10^{-4}$ ($\lamb_k^{\rm SF}$ and $Z_{C,k}^{\rm SF}$). $\lamb$ and $Z_C$ stand for $\lamb_{k=0}$ and $Z_{C,k=0}$, respectively. }
\label{fig_flow}
\end{figure}

Figure~\ref{fig_flow} shows the flow of the coupling constants $Z_{C,k}$ and $\lamb_k$ in the zero-temperature superfluid phase near a QCP. Exactly at the QCP ($\delta\mu=0$), one can clearly distinguish two regimes: i) a high-energy (or short-distance) regime $k\gtrsim k_x$ where lattice effects are important and the dimensionless coupling constant~\cite{Rancon11b} 
\begin{equation}
\tlamb_k = \frac{k}{Z_{C,k}Z_{A,k}t} \lamb_k 
\end{equation}
is large, ii) a weak-coupling (``Bogoliubov'') regime $k\lesssim k_x$ where $\tlamb_k\ll 1$ and the flow is governed by the Gaussian fixed point $\tlamb=0$: $\lamb_k$, $Z_{C,k}$ and $Z_{A,k}$ are then nearly equal to their fixed-point values [Eqs.~(\ref{maeff})] while $\tlamb_k\propto k$ vanishes in agreement with its scaling dimension $[\tlamb_k]=4-d-z=-1$ at the Gaussian fixed point ($d=3$ and $z=2$). In the momentum regime $|\q|\lesssim k_x$, the quasi-particles with mass $m^*$ and scattering length $a^*$ introduced in Sec.~\ref{subsec_qcp} are well defined and the physics becomes universal. The crossover scale $k_x$ between the two regimes is typically of the order of $\Lamb=\sqrt{6}$ ($k_x^{-1}$ is equal to a few lattice spacings). 

Away from the QCP, chemical potential and temperature introduce two new momentum scales, the ``healing'' scale 
\begin{equation}
k_h = \sqrt{2m^*|\delta\mu|} 
\end{equation}
and the thermal scale
\begin{equation}
k_T = \sqrt{2m^*T} . 
\end{equation}
Universality requires $k_h,k_T\ll k_x$. Since $k_x\sim a^{-1}\sim 1$ (except close to the tip of the Mott lobe) these conditions can be rewritten as
\begin{align}
\sqrt{m^*a^*{}^2|\delta\mu|}& \ll 1 , \label{univ} \\ 
\sqrt{m^*a^*{}^2T} &\ll 1 .
\end{align}
In the low-energy limit the system behaves as a gas of weakly-interacting quasi-particles if the dimensionless coupling constant 
\begin{equation}
\tlamb_{k_h} = \frac{k_h}{Z_{C,k_h}Z_{A,k_h}t} \lamb_{k_h} \simeq 8\pi k_h a^* 
\label{lambh}
\end{equation}
is small. The last result in~(\ref{lambh}) is obtained using $k_h\ll k_x$, which allows us to approximate $Z_{C,k_h}$, $Z_{A,k_h}$ and $\lamb_{k_h}$ by their $k=0$ values. Since $k_x\sim a^{-1}\sim 1$, universality ($k_h\ll k_x$) implies weak coupling ($k_ha^*\ll 1$). Using Eq.~(\ref{dnbar}), the weak-coupling/universality condition $k_h a^*\ll 1$ [Eq.~(\ref{univ})] can be rewritten as 
\begin{equation}
\sqrt{|\bar n-\bar n_c|a^*{}^3} \ll 1 .
\label{dnbar1}
\end{equation}
Equation~(\ref{dnbar1}) is similar to the usual condition for a boson gas to be dilute except that it involves the excess density of particles (or holes) $|\bar n-\bar n_c|$ (with respect to the commensurate density of the Mott insulator) rather than the full density $\bar n$ of the fluid. 

\begin{figure}
\centerline{\includegraphics[width=7cm]{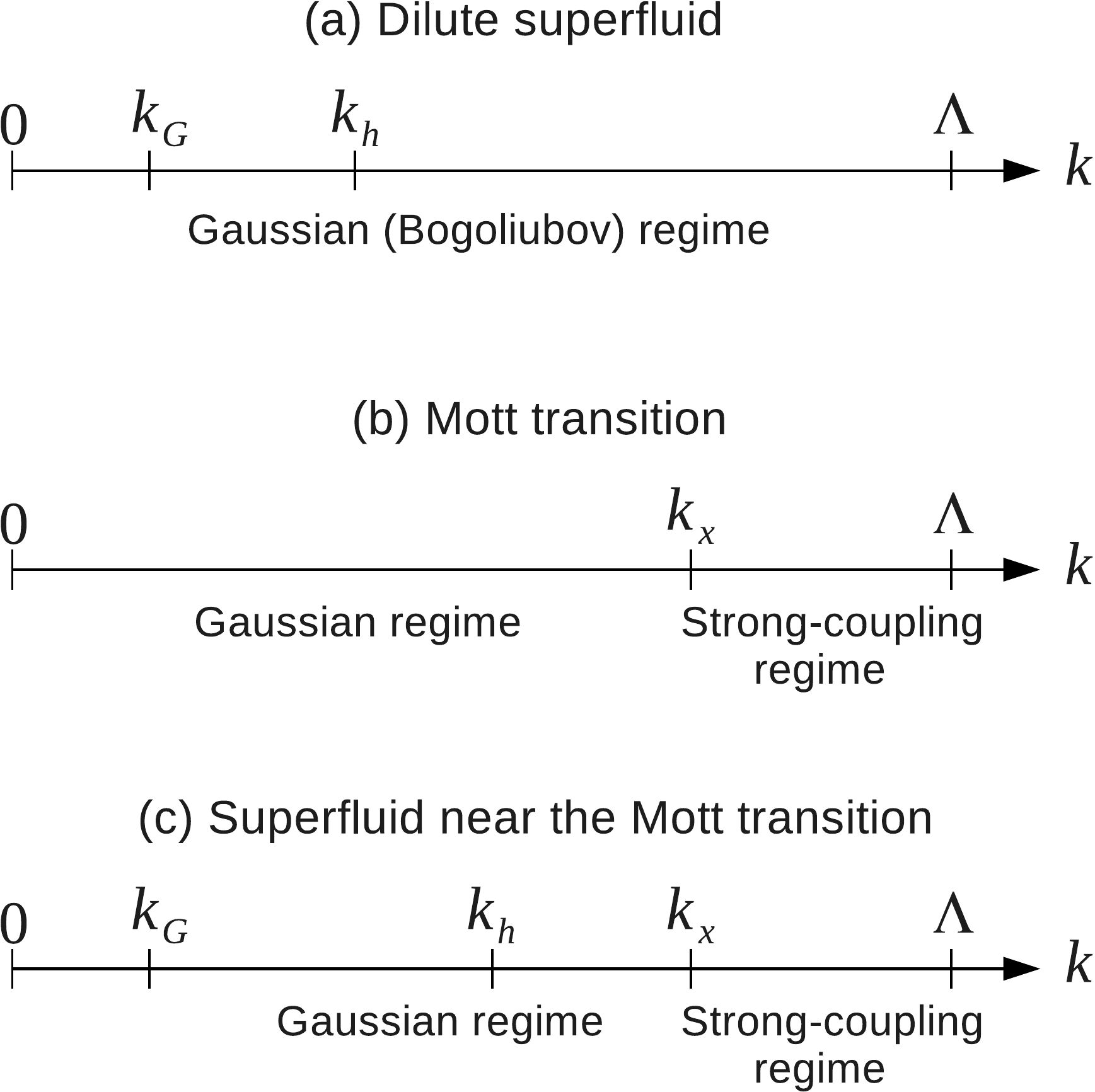}}
\caption{Characteristic momentum scales in a dilute superfluid, at the generic Mott transition ($\mu=\mu_c$), and in the superfluid phase near the Mott transition. (In the dilute superfluid, $\Lamb^{-1}$ is of the order of the scattering length $a$, while it is of the order of the inverse lattice spacing in the Bose-Hubbard model.)}
\label{fig_flow1} 
\end{figure}

For $k\lesssim k_h$, $\lamb_k$ and $Z_{C,k}$ depart from their fixed-point values at $\delta\mu=0$ (Fig.~\ref{fig_flow}) and vanish logarithmically below a ``Ginzburg'' momentum scale $k_G$ which is exponentially small at weak coupling ($\tlamb_{k_h}\ll 1$). In a dilute Bose gas, the Ginzburg scale manifests itself by the appearance of infrared divergences in the perturbation theory about the Bogoliubov approximation. Although these divergences cancel out for thermodynamic quantities, they do have a physical origin: they result from the coupling between longitudinal and transverse (phase) fluctuations and lead to a divergence of the longitudinal susceptibility -- a general phenomenon in systems with a broken continuous symmetry~\cite{note4}. For $k\sim k_G$, the RG flow crosses over to a ``Goldstone'' regime where the physics is dominated by phase fluctuations. We refer to Refs.~\cite{Dupuis09a,Dupuis09b,Sinner10} for a detailed discussion of the infrared behavior in the superfluid phase. The various 
regimes of the RG flow are summarized in Fig.~\ref{fig_flow1}.

\section{Conclusion}

We have presented a detailed study of the thermodynamics of a Bose gas near the generic (density-driven) Mott transition in the framework of the Bose-Hubbard model. In the critical regime, the physics is governed by weakly interacting quasi-particles with quasi-particle weight $\Zqp$, effective mass $m^*$ and ``scattering length'' $a^*$. Thermodynamic quantities can be expressed using the universal scaling functions of the dilute Bose gas universality class. They are independent of the quasi-particle weight and the only nonuniversal parameters entering the scaling functions are $m^*$ and $a^*$.  A notable exception is the condensate density $n_0$, which is proportional to $\Zqp$, thus allowing us to determine the quasi-particle weight from a thermodynamic measurement once $m^*$ and $a^*$ are known. The NPRG enables to compute $\Zqp$, $m^*$ and $a^*$ as a function of $t/U$. We find that the strong-coupling RPA, although rather inaccurate to determine the phase diagram, gives reliable estimates of $\Zqp$, $m^*
$ and $a^*$ as a function of $t/t_c$ (with $t_c$ the value of the hopping amplitude at the tip of the Mott lobe). 

The thermodynamics of a two-dimensional Bose gas in an optical lattice has recently been measured near the superfluid-vacuum transition~\cite{Zhang12}. A similar experiment in a three-dimensional gas would allow us to test the universality class of the generic three-dimensional Mott transition and the predictions of the NPRG approach regarding the values of $\Zqp$, $m^*$ and $a^*$. A measurement of the temperature dependence of the pressure in the quantum critical regime at $\delta\mu=0$ [Eq.~(\ref{Pmuc})] would directly provide us with the value of the effective mass $m^*$. Quite interestingly, $m^*$ strongly varies with both the ratio $t/U$ and the commensurate value of the density in the Mott insulator [Eq.~(\ref{tzero})]. Measuring the scattering length $a^*$ and the quasi-particle weight $\Zqp$ is more challenging as it would require to reach temperatures much smaller than the crossover temperature $T\sim 2t$ below which the thermodynamics becomes universal, which is not possible yet in actual 
experiments.

\begin{acknowledgments}
We would like to thank B. Capogrosso-Sansone for providing us with the QMC data~\cite{Capogrosso07} shown in Fig.~\ref{fig_ma_eff}.  
\end{acknowledgments} 

\appendix

\section{Perturbative calculation of scaling functions} 
\label{sec_scaling}

In this Appendix, we briefly review the perturbative calculation of the universal scaling functions of the three-dimensional dilute Bose gas universality class (Sec.~\ref{subsec_limitcases}).  

At low temperatures and positive chemical potential, the scaling functions can be obtained from a one-loop calculation (Bogoliubov theory). To organize the loop expansion~\cite{[{See, {\it e.g., }}] Zinn_book}, we introduce a parameter $l$ (which will eventually be set to 1) and consider the partition function 
\begin{equation}
Z[J^*,J] = \int\calD[\psi^*,\psi]\, e^{-l\left( S[\psi^*,\psi] - \inttau \int d^3 r (J^*\psi+\cc)\right)} 
\end{equation}
in the presence of a complex external source $J$. The superfluid order parameter is defined by
\begin{equation}
\phi(\r,\tau) = \frac{1}{l} \frac{\delta\ln Z[J^*,J]}{\delta J^*(\r,\tau)}, \quad
\phi(\r,\tau)^* = \frac{1}{l} \frac{\delta\ln Z[J^*,J]}{\delta J(\r,\tau)}. 
\end{equation}
We now introduce the effective action
\begin{equation}
\Gamma[\phi^*,\phi] = - \frac{1}{l} \ln Z[J^*,J] + \inttau \int d^3r (J^*\phi+\cc) ,
\end{equation}
defined as the Legendre transform of the thermodynamic potential $-l^{-1}\ln Z[J^*,J]$. The loop expansion is an expansion in $1/l$. To one-loop order, 
\begin{equation}
\Gamma[\phi^*,\phi] = S[\phi^*,\phi] + \frac{1}{2l} \Tr\ln\left(-\calG_c^{-1}[\phi^*,\phi]\right) + \calO(l^{-2}) , 
\label{gam1loop}
\end{equation}
where $\calG_c[\phi^*,\phi]$ is the classical propagator 
\begin{equation}
\calG_c[x,x';\phi^*,\phi] = -\left(
\begin{array}{cc}
\frac{\delta^{(2)}S}{\delta\phi^*(x)\delta\phi(x')} & \frac{\delta^{(2)}S}{\delta\phi^*(x)\delta\phi^*(x')} \\
\frac{\delta^{(2)}S}{\delta\phi(x)\delta\phi(x')} & \frac{\delta^{(2)}S}{\delta\phi(x)\delta\phi^*(x')}
\end{array}
\right) 
\end{equation}
(we use the notation $x=(\r,\tau)$ and $x'=(\r',\tau')$). 

All thermodynamic quantities can be obtained from the effective potential defined by 
\begin{equation}
V(n) = \frac{1}{\beta V} \Gamma[\phi^*,\phi] \Bigl|_{\phi\,\const} 
\end{equation}
($V$ denotes the volume of the system), where $\phi$ is a constant (uniform and time-independent) field. The U(1) symmetry of the action~(\ref{action1}) implies that $V(n)$ is a function of the condensate density $n=|\phi|^2$. The minimum of the effective potential determines the condensate density $n_0$ and the thermodynamic potential $V_0=V(n_0)$ per unit volume in the equilibrium state. The pressure is then given by 
\begin{equation}
P(\mu,T) = - V_0 .
\end{equation}
To compute the effective potential to one-loop order, we need to evaluate the trace in Eq.~(\ref{gam1loop}) with the classical propagator $\calG_c$ evaluated in a constant field,
\begin{equation}
\calG_c^{-1}(q;\phi^*,\phi) = \left( 
\begin{array}{cc}
i\wn-\xi_\q-2g|\phi|^2 & -g\phi^2 \\
-g\phi^*{}^2 & -i\wn -\xi_{-\q}-2g|\phi|^2 
\end{array}
\right) , 
\end{equation}
where $\wn=n 2\pi/\beta$ ($n$ integer) is a bosonic Matsubara frequency, $\xi_\q=\eps_\q-\mu$ and $\eps_\q=\q^2/2m$.

\subsection{Zero temperature} 

Let us first consider the zero-temperature limit. Performing the trace over Matsubara frequencies~\cite{note3}, we obtain 
\begin{equation}
V(n) = -\mu n + \frac{g}{2} n^2 + \frac{1}{2l} \int_\q (E_\q-\xi_\q-2gn) + \calO(l^{-2}) , 
\end{equation}
where $\int_\q=\int d^3q/(2\pi)^3$ and 
\begin{equation}
E_\q = [(\xi_\q+2gn)^2-(gn)^2]^{1/2} .
\label{Eq}
\end{equation}
The condensate density $n_0$ in the equilibrium state is obtained from $V'(n_0)=0$, \ie
\begin{equation}
n_0 = \frac{\mu}{g} - \frac{1}{2l} \int_\q \left[\frac{1}{E_\q}(2\xi_\q+3gn_0)-2\right] + \calO(l^{-2}),
\label{n0app} 
\end{equation}
where $E_\q$ is defined by~(\ref{Eq}) with $n=n_0$. 

Setting $n_0=\mu/g$ in the $\calO(1/l)$ term, we obtain the effective potential
\begin{equation}
V_0 = -\frac{\mu^2}{2g} + \frac{1}{2l} \int_\q (E_\q-\eps_\q-\mu) + \calO(l^{-2}) ,
\end{equation}
where $E_\q=\sqrt{\eps_\q(\eps_\q+2\mu)}$. To eliminate the dependence on the momentum cutoff $\Lambda$, we introduce the $s$-wave scattering length $a$,
\begin{equation}
\frac{m}{4\pi a} = \frac{1}{g} + \int_\q \frac{\Theta(\Lambda-|\q|)}{2\eps_\q} ,
\label{aapp}
\end{equation}
and rewrite $V_0$ as 
\begin{align}
V_0 ={}& - \frac{m\mu^2}{8\pi a} + \frac{\mu^2}{4} \int_\q \left(\frac{1}{\eps_\q}-\frac{1}{E_\q}\right) \nonumber \\ & + \half \int_\q \left(E_\q-\eps_\q-\mu + \frac{\mu^2}{2E_\q} \right) 
\label{V1}
\end{align}
setting $l=1$. The sums over $\q$ are now convergent and we can take the infinite cutoff limit $\Lambda\to\infty$ with $a$ fixed. Using 
\begin{equation}
\begin{gathered}
\int_\q \left(\frac{1}{\eps_\q}-\frac{1}{E_\q}\right) = \frac{2}{\pi^2} m^{3/2}\mu^{1/2} , \\
\int_\q \left(E_\q-\eps_\q-\mu + \frac{\mu^2}{2E_\q} \right) = \frac{m^{3/2}\mu^{5/2}}{15\pi^2} ,
\end{gathered}
\end{equation}
we finally obtained 
\begin{equation}
V_0 = - \frac{m\mu^2}{8\pi a} \left(1-\frac{64}{15\pi} \sqrt{ma^2\mu}\right) 
\label{V2}
\end{equation}
and in turn Eq.~\eqref{P1}. Equations~(\ref{n1},\ref{kappa1}) are then deduced from $\bar n=-\partial V_0/\partial\mu$ and $\kappa=\partial\bar n/\partial\mu$. 

Similarly, from~(\ref{n0app},\ref{aapp}) we deduce 
\begin{equation}
n_0 = \frac{m\mu}{4\pi a} - \frac{\mu}{2} \int_\q \left(\frac{1}{\eps_\q}-\frac{1}{E_\q}\right) + \int_\q \left(1 - \frac{\eps_\q+\mu}{E_\q}\right) 
\end{equation}
to one-loop order (setting $l=1$), which leads to Eq.~(\ref{n0_1}) using 
\begin{equation}
\int_\q \left(1 - \frac{\eps_\q+\mu}{E_\q}\right) = -\frac{2}{3\pi^2} (m\mu)^{3/2} .
\end{equation}

The superfluid density $n_s$ is defined by the variation 
\begin{equation}
\delta\Gamma = \beta \frac{n_s}{2m} \int d^3r (\nablabf\theta)^2 
\end{equation}
of the effective action when the superfluid order parameter $\phi(\r)=\sqrt{n_0}e^{i\theta(\r)}$ acquires a phase slowly-varying in space. For a dilute Bose gas, Galilean invariance implies that $n_s$ is equal to the fluid density, {\it i.e.} $n_s(\mu,0) = \bar n(\mu,0)$. To leading order, the sound mode velocity $c=\sqrt{\rho_s/\kappa}=\sqrt{n_s/m\kappa}$ is equal to $\sqrt{\mu/m}$, in agreement with the small-$\q$ behavior of $E_\q=\sqrt{\eps_\q(\eps_\q+\mu)}$. 

\subsection{Finite temperature} 

Similarly, we can compute the pressure $P(0,T)$ at vanishing chemical potential. For $\mu=0$, the condensate density $n_0=0$ in the equilibrium state. The effective potential $V_0$ to one-loop order is then simply given by the non-interacting result,
\begin{align}
V_0 &= \frac{1}{\beta} \int_\q \ln\left(1-e^{-\beta\eps_\q}\right) \nonumber \\ 
&= - \zeta(5/2) \left(\frac{m}{2\pi}\right)^{3/2} T^{5/2} , 
\label{V3}
\end{align}
where $\zeta(x)$ is the Riemann zeta function.

\subsection{Transition line}

The one-loop approximation fails near the superfluid transition temperature $T_c$~\cite{Baym77}. The transition temperature can nevertheless be determined from a perturbative approach in the normal state, by considering the self-consistent one-loop self-energy correction (self-consistent Hartree-Fock approximation),
\begin{align}
\Sigma &= -\frac{2g}{\beta}\sum_{\wn} \int_\q \frac{e^{i\wn 0^+}}{i\wn-\xi_\q-\Sigma} \nonumber \\ 
&= 2g \int_\q n_B(\eps_\q+\Sigma-\mu) ,
\end{align}
where $n_B(x)=(e^{\beta x}-1)^{-1}$ is the Bose-Einstein distribution function. The transition occurs when the renormalized chemical $\mu-\Sigma$ vanishes, 
\begin{equation}
\mu = 2g \int_\q n_B(\eps_\q) = 2g \zeta(3/2)\left(\frac{mT_c}{2\pi}\right)^{3/2} .
\end{equation}
To lowest order $g=4\pi a/m$, which gives Eq.~\eqref{muc}.

\section{Strong-coupling RPA}
\label{sec_rpa}

In the strong-coupling RPA~\cite{Sheshadri93,Oosten01,Sengupta05,Ohashi06,Menotti08}, the effective action is given by~(\ref{GamLambda}). This implies that in the Mott insulating phase, the single-particle propagator takes the form
\begin{equation}
G(\q,i\w) = \frac{G_{\rm loc}(i\w)}{1-t_\q G_{\rm loc}(i\w)} , 
\end{equation}
where 
\begin{equation}
G_{\rm loc}(i\w) = \frac{\bar n+1}{i\w+\mu-U\bar n} - \frac{\bar n}{i\w+\mu-U(\bar n-1)}
\end{equation}
is the local propagator. $\bar n$ denotes the mean (integer) number of bosons per site. The instability of the Mott insulator is signaled by the appearance of a pole in the propagator, i.e. $1-t_{\q=0}G_{\rm loc}(i\w=0)=0$, which reproduces the mean-field phase diagram~\cite{Fisher89}. The value of the hopping amplitude at the tip of the Mott lobe is given by
\begin{equation}
\frac{2dt_c}{U} = 2\bar n+1 -2 \sqrt{\bar n^2+\bar n} . 
\end{equation}
On the transition line, we obtain 
\begin{equation}
G(\q,i\w) = \frac{\Zqp}{i\w-\q^2/2m^*}  
\end{equation}
for $\q,\w\to 0$, with 
\begin{equation}
\Zqp = \frac{m}{m^*} = \left| \frac{G_{\rm loc}(0)}{2dt G'_{\rm loc}(0)} \right| , 
\label{rpa1} 
\end{equation}
where $G'_{\rm loc}(i\w)=\partial_{i\w}G_{\rm loc}(i\w)$. As in Sec.~\ref{subsec_qcp}, we have performed a particle-hole transformation when we consider the lower branch of the transition line (hence the absolute value in Eq.~(\ref{rpa1})). 

In the limit $t\to 0$, Eq.~(\ref{rpa1}) leads to~(\ref{tzero}), a result which can be understood as follows. Let us add a particle at site $\r$ to a Mott insulator with $\bar n$ particles per site. In the limit $t\to 0$, the only possible dynamics is due to the motion of the additional particle. Hopping of this particle between sites $\r$ and $\r'$ involves the matrix element 
\begin{equation}
\bra{\bar n ;\r}\otimes \bra{\bar n+1;\r'} t \hat\psi^\dagger_{\r'} \hat\psi_\r \ket{\bar n+1;\r} \otimes \ket{\bar n;\r'} = t (\bar n+1)  
\end{equation} 
if we denote by $\otimes_{\r_i}\ket{n_i;\r_i}$ the state with $n_i$ particles at site $\r_i$. The ``particle'' eigenstates are therefore plane wave states,
\begin{equation}
\ket{\q} = \frac{1}{\sqrt{N}} \sum_\r e^{i\q\cdot\r} \ket{\bar n+1;\r} \otimes_{\r'\neq \r} \ket{\bar n;\r'} ,
\end{equation}
with a dispersion law $\lamb_\q=\lamb_0 -2t(\bar n+1)\sum_{i=1}^d \cos q_i$ (with $\lamb_0$ a constant which takes the value $2dt(\bar n+1)$ at the quantum critical point), which leads to an effective mass $m^*/m=1/(\bar n+1)$. The single-particle propagator reads
\begin{equation} 
G(\q,i\w) = \frac{|\bra{\q} \hat\psi^\dagger(\q)\ket{0}|^2}{i\w-\lamb_\q} , 
\end{equation}
where $\ket{0}=\otimes_\r\ket{\bar n;\r}$ denotes the ground state of the Mott insulator without the additional particle (in the limit $t\to 0$). We deduce the quasi-particle weight
\begin{equation}
\Zqp = |\bra{\q} \hat\psi^\dagger(\q)\ket{0}|^2 = \bar n+1 .  
\end{equation}
A similar reasoning for the motion of a hole leads to $\Zqp=m/m^*=\bar n$.

\section{Ward identity $Z_\mu=Z_C$} 
\label{sec_Zmu}

To lowest order in $\delta\mu$, the effective action~(\ref{gamc}) at the QCP is modified by 
\begin{equation}
\delta\Gamma[\phi^*,\phi] = -  Z_\mu \delta\mu\inttau \int d^3r \phi^*\phi .
\end{equation}
This implies that the effective potential is given by 
\begin{equation}
V(\mu,n) = V(\mu_c,n) - Z_\mu n \delta\mu . 
\end{equation}
The invariance of the action $S$ in the local (time-dependent) gauge transformation 
\begin{equation}
\psi_\r\to\psi_\r e^{i\alpha}, \quad \psi^*_\r\to\psi^*_\r e^{-i\alpha}, \quad \mu\to\mu+i\dtau\alpha
\label{localgauge} 
\end{equation}
implies that 
$Z_C\equiv Z_C(\mu_c)$ satisfies the Ward identity~\cite{Rancon11b}
\begin{equation}
Z_C = - \frac{\partial^2 V}{\partial\mu\partial n} \biggl|_{\mu_c,n=0} , 
\label{wardid} 
\end{equation} 
which gives 
\begin{equation} 
Z_C = Z_\mu . 
\end{equation}

%\bibliography{/users/lptl/dupuis/publi/BIB/nprg.bib,/users/lptl/dupuis/publi/BIB/bosons.bib,/users/lptl/dupuis/publi/BIB/book.bib,/users/lptl/dupuis/publi/BIB/notes.bib,/users/lptl/dupuis/publi/BIB/stat_phys.bib,/users/lptl/dupuis/publi/BIB/bcsbec.bib,/users/lptl/dupuis/publi/BIB/cold_atoms.bib}

\begin{thebibliography}{38}
\bibitem [{not({\natexlab{b}})}]{note6}%
  \BibitemOpen
  \href@noop {} {} \bibinfo {note} {The NPRG result depends
  on the approximation used to solve the flow equations. We quote here the best
  result: see Refs.~\cite{Rancon11a,Rancon11b} for a detailed
  discussion.}\BibitemShut {Stop}%
\end{thebibliography}


\end{document}